\documentclass[11pt]{article}
\usepackage{graphicx}
\usepackage{psfrag}
\usepackage{amsmath}
\usepackage[latin1]{inputenc}
\usepackage[T1]{fontenc} 
\usepackage[french]{babel}
\def\at{\string@}
\def\t2#1{\underline{\underline{#1}}}

\def\stav#1{\langle #1 \rangle}

\def\th\lim{\displaystyle \lim_{th}}

\newcommand{\be}{\begin{equation}}
\newcommand{\ee}{\end{equation}}

\newcommand{\bi}{\begin{itemize}}
\newcommand{\ei}{\end{itemize}}
\newcommand{\bc}{\begin{center}}
\newcommand{\ec}{\end{center}}
\newcommand{\im}{\item}
\newcommand{\bn}{\begin{enumerate}}
\newcommand{\en}{\end{enumerate}}
\def\t2#1{\underline{\underline{#1}}}

\def\stav#1{\langle #1 \rangle}
\renewcommand{\author}[1]{
\begin{center}
{\bf #1}
\end{center}
\par
}
\renewcommand{\title}[1]{\begin{center}
    {\large{\bf #1}}
     \end{center}
\medskip
} 

\begin{document}
\title{  
SIMULATION NUMÉRIQUE DISCRÈTE
ET COMPORTEMENT MÉCANIQUE DES MATÉRIAUX GRANULAIRES}
\author{
Jean-Noël ROUX et François CHEVOIR}
\bc
Laboratoire des Matériaux et des Structures du Génie Civil, \\
Unité Mixte de Recherche LCPC-ENPC-CNRS, Institut Navier \\
2, allée Kepler, Cité Descartes, 77420 Champs-sur-Marne \\
\ec
\bc
\begin{minipage}{16cm}
{\small
   \parindent=0pt
    {\bf R\'ESUM\'E : }
Complémentaire des expériences de laboratoire, la simulation numérique discrète appliquée aux matériaux granulaires 
donne accès à la microstructure à l'échelle des grains et des contacts, et permet de comprendre l'origine microscopique 
des comportements mécaniques macroscopiques. Nous introduisons d'abord les différentes méthodes de  simulation discrète, 
considérées comme outils d'expérimentation numérique, en relation avec les modèles mécaniques des contacts entre grains. 
Nous insistons alors sur certains aspects importants de l'approche numérique discrète dans l'étude des comportements mécaniques : 
la représentativité des échantillons, l'utilisation de l'analyse dimensionnelle pour identifier les paramètres pertinents. 
Dans un second temps, nous décrivons l'application de cette démarche à deux grandes classes de comportement mécanique 
des matériaux granulaires : la déformation quasi-statique de matériaux granulaires solides, pour laquelle on souligne 
l'importance de la géométrie de l'assemblage; les lois de frottement et de dilatance qui régissent les écoulements denses, 
confinés ou à surface libre.
\par
}
\end{minipage}
\ec
\section{Introduction.}

Nous nous intéressons ici au comportement mécanique des matériaux granulaires,
que l'on rencontre dans le génie civil ou l'environnement, 
tels que les poudres, les sables, ou les granulats [1-7].
Ils sont constitués d'un grand nombre de grains macroscopiques qui interagissent 
localement au niveau de leurs contacts, par élasticité, frottement, collisions,
forces interfaciales, quelquefois par l'intermédiaire 
d'un autre corps sous forme liquide ou solide. 

Selon les conditions, ces matériaux présentent des comportements mécaniques variés,
qui les apparentent aux solides élastoplastiques pour 
des sables en régime quasi-statique (domaine de la mécanique des sols) [8],
aux fluides visqueux, viscoplastiques ou pâteux avec seuil 
lorsque l'on provoque un écoulement (comme lors de manutentions de granulats,
ou dans les problèmes d'écoulements sur des pentes 
en milieu naturel) [9], voire aux gaz denses sous forte agitation [10].
Pour le comportement rhéologique du matériau, et pour les conditions 
aux interfaces, on ne dispose pas toujours de lois constitutives adéquates
au niveau macroscopique. Il est naturel de chercher à formuler ou 
à améliorer les lois rhéologiques, en se fondant sur leurs origines physiques
à l'échelle des grains et de leurs interactions. 
L'étude micromécanique des matériaux granulaires est un champ d'investigation
récent, dans lequel l'usage des simulations numériques discrètes 
est venu  compléter les expériences physiques sur des matériaux modèles [11]. 

On a affaire à un modèle de matériau granulaire si on peut considérer que les interactions entre les grains, 
objets solides en général, sont localisées dans des régions de contact très petites devant leurs diamètres. 
Cette hypothèse garantit la possibilité de décrire la cinématique de l'ensemble comme celle d'une collection de corps rigides, 
avec un nombre fini de degrés de liberté. Ceci exclut les effets hydrodynamiques à longue portée qui interviennent dans les suspensions 
(où il est nécessaire de prendre un compte un champ de vitesse dans le fluide) [12]
mais on peut éventuellement intégrer dans cette définition 
les interactions visqueuses entre grains proches voisins qui dominent dans les pâtes granulaires [9]
ou encore dans les enrobés bitumineux [13]. 
Enfin, dans un matériau réel, les grains sont en général de forme quelconque, de surface irrégulière, 
et présentent très souvent une forte polydispersité [14].

Dès lors que sont précisés le modèle d'interaction entre grains ainsi que les sollicitations appliquées,
on conçoit que les matériaux granulaires 
se prêtent bien à la simulation numérique. Celle-ci (que l'on dit discrète parce que l'on
travaille avec un nombre fini de degrés de liberté) 
peut fournir beaucoup d'informations inaccessibles expérimentalement, puisque l'on calcule
toutes les positions, les vitesses et les efforts. 
De plus elle permet de varier à loisir les paramètres mécaniques des grains
et les sollicitations, donc de multiplier les « expériences numériques ».

Les temps de calcul nécessaires à l'obtention de résultats significatifs dépendent bien sûr de la taille du système (nombre de grains), 
du comportement étudié (il est évidemment plus coûteux de simuler les phénomènes lents), des lois d'interaction 
(qui donnent lieu à des calculs plus ou moins complexes). Si l'on souhaite avoir un ordre de grandeur sommaire, 
on peut noter que l'on est actuellement capable, au moyen d'un ordinateur personnel
ou d'une station de travail, de simuler une évolution importante, 
avec de notables changements de configuration,  d'un système de quelques milliers de grains en quelques heures de temps CPU. 
(Par évolution significative, on entend par exemple la déformation jusqu'au palier plastique en régime quasi-statique, 
ou bien l'étude d'un écoulement stationnaire, ou encore le phénomène de blocage). Certaines équipes disposant d'outils de calcul parallèle 
simulent aujourd'hui des échantillons granulaires composés de plusieurs centaines de milliers de grains [15, 16].
	
La partie 2 présente d'abord différents modèles mécaniques de contacts intergranulaires, et classifie les diverses méthodes de simulation. 
L'exposé sera limité au cas des grains circulaires en deux dimensions ou sphériques en trois dimensions.
Les méthodes restent essentiellement 
les mêmes pour des grains de forme plus compliquée (polyèdres [17-20],
ellipsoïdes [21]) mais les interactions sont parfois moins bien
connues et la mise en {\oe}uvre est plus lourde. La polydispersité n'est pas
une difficulté en soi, sinon pour le nombre de grains à prendre en
compte et donc le temps de calcul. Nous décrirons dans la suite des applications
à des systèmes de faible étendue granulométrique. Nous montrerons en
revanche  
que l'on est capable de traiter des modèles de contact variés, et des
échantillons numériques soumis à des sollicitations mécaniques très
diverses.  

La partie 3 présente quelques résultats obtenus pour la déformation
quasi-statique d'assemblages de grains sphériques monodisperses en  
3 dimensions, et la partie 4 est une étude des lois d'écoulement de
matériaux modèles bidimensionnels constitués de disques. 

\section{Généralités sur la simulation numérique discrète des matériaux granulaires}

\subsection{Interaction entre grains : paramètres des modèles }
 Pour un modèle mécanique de matériau granulaire, les interactions,
 limitées à des régions de contact de taille très faible  
devant celle des grains,  sont binaires et s'expriment par une loi de
 contact qui relie une force ponctuelle à la surface des deux  
objets en contact à l'histoire de leurs mouvements. 

Le comportement mécanique des contacts est complexe et souvent mal
connu, car il est lié à des détails fins de la physique des surfaces  
qu'il est vain d'espérer contrôler dans les matériaux du génie
civil. On est amené en pratique, pour les besoins du calcul,  
à adopter des lois de contact relativement simplifiées et robustes,
dont certains ingrédients sont assez bien connus, et d'autres beaucoup
moins.  
Il est donc important d'évaluer les conséquences, sur les
comportements mécaniques que l'on souhaite étudier, des paramètres que
l'on  
introduit dans les modèles de contact [22].

\begin{figure}[!htbp]
\bc
\includegraphics[width=16cm]{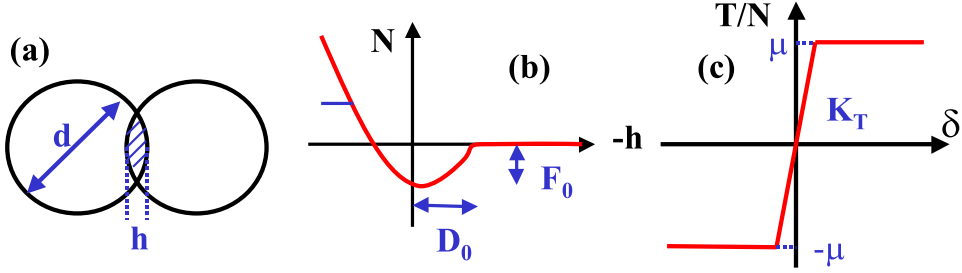}
\caption{Modèle de contact (dynamique moléculaire). 
(a) Déformation de la zone de contact ; 
(b) Modèle pour la force normale ; 
(c) Modèle pour la force tangentielle.
\label{fig1}
}
\ec
\end{figure}

On considère deux grains identiques (disques ou sphères) de diamètre $d$
et masse $m$ (Fig.~\ref{fig1}).  Selon le comportement ou le phénomène étudié, 
on peut choisir de modéliser les déformations du contact, ou bien
traiter les grains comme parfaitement rigides et indéformables.  
\subsubsection{Grains élastiques frottants}
L'élasticité du contact est relativement bien connue dans le cas
d'objets réguliers sans arêtes, comme les sphères [23].  
La loi de Hertz relie alors, pour des grains sphériques de diamètre d
constitués d'un matériau élastique de module d'Young $E$ 
et de coefficient de Poisson $\nu$, la force normale $F_N$  à la
déflexion normale $h$  des surfaces en contact (« interpénétration »
apparente) : 
\be
F_N=\frac{Ed^{1/2}}{3(1-\nu ^2)}h^{3/2}
\label{eqn:hertz}
\ee
La variation de la partie élastique de la réaction tangentielle $F_T$
avec le déplacement relatif tangentiel $\delta$ [23], 
du fait du critère de Coulomb écrit au niveau local du vecteur-contrainte dans la petite surface de contact, 
ne peut en principe s'écrire que de façon incrémentale, et donne des lois fort compliquées. On est souvent 
(comme dans les résultats donnés au §3) amené à les simplifier, en gardant la raideur tangentielle,
\be
\frac{dF_T}{d\delta}=\frac{Ed^{1/2}}{(2-\nu)(1+\nu)}h^{1/2}
\label{eqn:mindlin}
\ee
indépendante de $\delta$  tant que la condition de Coulomb 
\be
\vert\vert {\bf F}_T \vert\vert \le \mu F_N
\label{eqn:coulomb}
\ee
(où apparaît le coefficient de frottement $\mu$) est satisfaite. Dans
le cas contraire, une fois incrémentée ${\bf F}_T $ dans le calcul, 
il faut la projeter sur le cône de Coulomb. Il faut de surcroît se
préoccuper de l'effet des variations simultanées de $F_N$  et de  ${\bf F}_T$, 
et transporter les forces de contact avec les mouvements d'ensemble et
relatifs (y compris roulements et pivotements) des deux objets.  
Pour ce faire on ne dispose pas en fait de loi établie pour tous les
cas, et on doit se contenter de règles praticables  
qui ne violent ni l'objectivité, ni le second principe de la thermodynamique [24].

Il arrive aussi très souvent que l'on ne cherche pas à décrire
précisément l'élasticité du contact, et que l'on choisisse  
plus simplement une loi linéaire unilatérale : 
\be
F_N=K_N h,\ \ \ \frac{dF_T}{d\delta}=K_T,
\label{eqn:linuni}
\ee
avec la condition de Coulomb~\eqref{eqn:coulomb}. Les raideurs normale
($K_N$) et tangentielle ($K_T$) peuvent être choisies pour respecter l'ordre de grandeur 
des déformations élastiques dans les contacts, ou bien on les
considère simplement comme une manière commode d'assurer  
l'impénétrabilité des grains et la mobilisation progressive du
frottement.
\subsubsection{Viscosité}

La partie visqueuse de la force de contact est le plus souvent prise linéaire dans la vitesse relative : 
\be
F_N^v=\alpha_N \frac{dh}{dt},\ \ \ F_T^v=\alpha_T \frac{d\delta}{dt},
\label{eqn:visco}
\ee
avec des coefficients d'amortissement dont l'origine physique n'est pas bien connue en général 
(pour une modélisation fondée sur la viscoélasticité du matériau solide constituant les grains, voir toutefois [25]). 
La modélisation d'un choc binaire frontal avec les relations~\eqref{eqn:linuni} et \eqref{eqn:visco}
définit un problème d'oscillateur harmonique avec amortissement linéaire, 
et fixe la durée d'une collision : 
\be
\tau_c=\pi\left[ \frac{2K_N}{m}-\left( \frac{\alpha_N}{m}\right)^2\right]^{-1/2}.
\label{eqn:tauc}
\ee
Lorsque l'on souhaite surtout, afin d'approcher rapidement des états d'équilibre que l'on veut étudier, dissiper efficacement l'énergie, 
on tend alors à choisir des coefficients d'amortissement « critiques ». Dans le cas hertzien~(\ref{eqn:hertz}), on peut éventuellement 
se référer à la raideur linéaire tangente. Notons enfin qu'après avoir introduit des termes visqueux dans la force de contact, 
on doit se demander si on les inclut dans les composantes $F_N$  et ${\bf F}_T$  qui doivent satisfaire l'inégalité de Coulomb (3). 
Dans les exemples traités aux §3 et §4, on a toujours mis en {\oe}uvre le critère de Coulomb pour les seules parties élastiques des forces, 
et choisi  $\alpha_T=0$. 

\subsubsection{Plasticité}

Une autre source possible de dissipation est la plasticité du  matériau constituant les grains. 
On trouvera dans [23] des modélisations du contact élastoplastique d'objets à surface régulière (telles que les billes métalliques), 
qui complètent les descriptions élastiques, et dans [26] un exemple de mise en oeuvre d'un modèle simplifié de plasticité des contacts. 
C'est un ingrédient dont nous ne traitons pas dans la suite de cet article.

\subsubsection{Cohésion}

Par ailleurs, on peut introduire dans la loi de contact un effet d'adhésion. Celui-ci entraîne en général 
l'introduction de deux nouveaux paramètres dans le modèle de contact, une force attractive maximale  $F_0$, 
et une portée de l'interaction attractive  $D_0$. L'adhésion peut provenir de la capillarité en présence d'un pont liquide (ménisque) 
[27, 28], ou de l'interaction directe entre surfaces solides, que nous discutons maintenant. 
Celle-ci est négligeable pour des particules millimétriques (grains de sable), mais importante aux échelles colloïdales, 
dans les matériaux plus finement divisés (poudres, argiles). Dans le cas de fines particules solides interagissant à distance par 
l'attraction de van der Waals,  $D_0$ est de l'ordre de quelques
nanomètres, tandis que le produit  $D_0F_0$ est voisin de l'énergie interfaciale  $\gamma$. 
On dispose en principe des modèles de Johnson-Kendall-Roberts (JKR) et de Deriaguine-Muller-Toporov (DMT) [29]. 
Le modèle JKR correspond à la limite dans laquelle la portée des forces d'adhésion est négligée devant la déflexion élastique 
du contact qu'elles provoquent. La limite opposée, dans laquelle l'attraction à distance est prise en compte mais la déformation des objets 
solides qu'elle entraîne est négligée, est décrite par le modèle DMT. Certaines approches [29, 30] permettent de décrire 
les situations intermédiaires entre JKR et DMT. Un paramètre sans dimension (parfois appelé paramètre de Tabor) décrit 
le passage entre ces deux limites, il est donné (à un facteur numérique près) par
 
\be
\varpi=\frac{1}{D_0}\left( \frac{\gamma^2 d}{E^2}\right)^{1/3}
\label{eqn:tabor}
\ee

et on montre aisément que la déflexion normale élastique $h_0$  du
contact due à une force normale égale à  $\gamma d$
(qui est l'ordre de grandeur de la force d'adhésion selon DMT, ou de la force de traction maximale dans tous les cas) 
vérifie d'après la loi de Hertz~(\ref{eqn:hertz})
\be
\frac{h_0}{D_0} \propto \varpi ^3.
\label{eqn:h0tabor}
\ee
Dans le cas de la cohésion capillaire [27, 28], l'attraction maximale
sera également d'ordre $\gamma d$, $\gamma$   
désignant alors la tension superficielle liquide-vapeur. La portée  $D_0$ étant liée au volume du ménisque, 
excèdera facilement $h_0$ et la situation s'apparentera donc au cas DMT.

En pratique, les simulations numériques ont jusqu'ici simplifié assez drastiquement les lois de contact avec 
adhésion plutôt que cherché à mettre en {\oe}uvre des modèles très précis (voir toutefois [31] pour un calcul de collision binaire 
avec le modèle JKR). Un modèle simple à un seul paramètre [32-34] utilisé au §4.2.2 dans les systèmes bidimensionnels consiste à ajouter, 
dans le modèle avec élasticité linéaire~(\ref{eqn:linuni}) une force attractive de portée nulle 
en dehors du contact et proportionnelle à la « surface » 
de recouvrement, soit
\be
F^C=-\gamma \sqrt{dh},
\label{eqn:attrac1}
\ee
avec $\gamma$ une énergie d'interface (Fig. 1). On a alors $h_0=d
(\gamma /K_N)^2$ pour l'interpénétration d'équilibre d'une paire de grains en contact. 
Ce modèle est plutôt du type JKR (puisque $D_=0$), mais avec une
résistance à la traction $F_0^C=\gamma^2 d /(4K_N)$  qui dépend de la raideur $K_N$  (contrairement au modèle JKR).

\subsubsection{Lubrification}

	En présence d'un fluide interstitiel visqueux, il devient indispensable (sauf dans la limite quasi-statique des 
très faibles vitesses) de prendre en compte les interactions hydrodynamiques entre grains. Si le matériau granulaire est très concentré 
(« pâte granulaire »), on peut penser que les interactions binaires entre grains voisins vont jouer un rôle dominant. 
On utilisera l'approximation de lubrification pour décrire les interactions visqueuses [23].

\subsubsection{Efforts liés au roulement ou au pivotement}

La description du contact comme ponctuel ou de très faible étendue (devant le diamètre $d$) entraîne que les moments de roulement ou de 
pivotement évalués en un point de la région de contact, produits des forces résultantes par un bras de levier très petit devant d, 
peuvent être négligés en général. Toutefois, avec des grains rugueux ou de forme irrégulière, la zone de contact, éventuellement non connexe,
 peut posséder une certaine étendue et transmettre un moment. Un modèle destiné à prendre en compte de tels effets, en dimension 2, 
a été proposé par Iwashita et Oda [35], en gardant par ailleurs des grains circulaires. Il introduit plusieurs paramètres supplémentaires, 
associé à l'élasticité et au seuil plastique en roulement.	

\subsubsection{Grains rigides}

On peut également choisir, en particulier dans les systèmes fortement agités, de ne pas modéliser la déformation des contacts, 
mais de traiter simplement de chocs instantanés. En effet, les déplacements associés aux déformations dans les contacts peuvent 
alors être négligés devant les mouvements des grains. Le modèle classique de choc avec « coefficients de restitution » énonce 
simplement que la vitesse relative (normale ou tangentielle) après le choc est égale à une certaine fraction (le coefficient de restitution, 
normal, $e_N$  , ou tangentiel, $e_T$ ) de l'opposé de la vitesse relative antérieure au choc. De telles modélisations ont reçu des confirmations 
expérimentales [36] dans le cas des collisions binaires de billes sphériques. Cependant, elles ne précisent pas comment décrire les 
contacts maintenus et les collisions multiples. En fait, les lois élastique (3) et visqueuse (4) ci-dessus se traduisent, 
dans une collision isolée, par un coefficient de restitution normal

\be
e_N=\exp\left[\frac{\pi\zeta}{2\sqrt{1-\zeta^2}}\right],
\label{eqn:restitn}
\ee
$e_N$  désignant le rapport du coefficient d'amortissement $\alpha_N$  à sa valeur critique. Les collisions tangentielles sont plus complexes 
(il peut y avoir  glissement pendant une fraction de la durée du contact, qui dépend de l'angle d'incidence, etc.). 
Enfin, les calculs de [25] donnent des coefficients de restitution normaux qui dépendent de la vitesse relative avant le choc.

\subsection{Méthodes de simulation numérique discrète}

Les simulations numériques discrètes de milieux granulaires [11, 37, 38] ont été initialement développées pour les 
évolutions quasi-statiques d'empilements granulaires denses. Elles ont ensuite été appliquées à des situations dynamiques 
(écoulement, vibration). Suivant que l'échelle microscopique liée aux déformations locales des particules est négligée ou pas, 
on parle de grains déformables ou de grains rigides. On distinguera aussi les méthodes de simulation dynamiques 
(dans lesquelles inertie et mécanismes de dissipation interviennent explicitement) et les méthodes statiques 
(recherche d'une succession d'états d'équilibre). 

\subsubsection{Dynamique moléculaire. Cas des grains déformables}

Les méthodes de dynamique moléculaires furent inventées à partir de la fin des années 50, dès l'apparition des premiers ordinateurs,  
pour simuler les liquides et les solides [39-41]. Les particules sont alors des atomes ou des molécules en interaction et en mouvement 
permanent, avec conservation de l'énergie. La simulation a pour but, en particulier, de retrouver leurs propriétés thermodynamiques et de transport 
à partir de l'échelle particulaire. Plusieurs décennies de développements, joints à la puissance des ordinateurs actuels, 
permettent aujourd'hui de simuler des fluides et des matériaux de plus en plus complexes (systèmes moléculaires, macromoléculaires, colloïdaux) [42].

Dans les années 80, la dynamique moléculaire fut adaptée aux assemblages de grains solides légèrement déformables, 
et souvent dans ce contexte rebaptisée méthode aux éléments discrets [43]. La démarche la plus répandue consiste à prendre 
en compte la déformation élastique des contacts, selon les lois (2) et (3), ou bien (4), avec la loi de frottement de Coulomb. 
L'algorithme consiste alors, à chaque pas de temps, d'abord à détecter les grains en contact, puis à calculer les forces de contact binaires, 
enfin à intégrer les relations fondamentales de la dynamique pour tous les grains de façon à modifier leur vitesse et leur position. 
Ceci nécessite une discrétisation en temps du système d'équations différentielles qui en résulte, selon l'un ou l'autre des schémas 
explicites décrits dans [39]. Le pas de temps approprié est alors typiquement une petite fraction de la durée d'une collision (6) 
(ou de la pseudo-période des oscillations amorties d'un contact). 

Cette méthode conduit en pratique à effectuer un très grand nombre d'itérations, chacune d'entre elles étant toutefois 
peu coûteuse en temps de calcul. Diverses techniques (listes de voisins, découpage en cellules...) [39, 41] permettent 
d'éviter la recherche des contacts parmi toutes les paires possibles ( avec n grains) et de calculer avec un nombre d'opérations 
élémentaires proportionnel au nombre de grains.

D'une grande souplesse d'utilisation, ce type de dynamique moléculaire peut donner lieu quelquefois à des calculs 
inutilement longs dans les cas où les échelles de temps et d'espace associées à la déformation d'un contact ne sont pas pertinentes.

\subsubsection{Méthodes pilotées par événements}

Pour éviter de tenir compte des petites échelles d'espace et de temps dans les interactions entre grains, 
on est amené à les considérer comme des solides rigides, impénétrables. La loi de contact est alors remplacée 
par une loi de choc qui dépend des trois paramètres mécaniques introduits au §2.1~: les coefficients de restitution normal $e_N$, 
tangentiel $e_N$ et de frottement $\mu$. Lors d'une collision binaire,
les vitesses des deux grains concernés sont immédiatement modifiées selon cette loi. 

Le mouvement des grains n'est plus défini par une équation différentielle ordinaire, mais se présente comme une séquence de 
collisions entre lesquelles les vitesses (en l'absence de force extérieure) restent constantes. 
On désigne sous le nom de méthode pilotée par événements («~event-driven~») la procédure qui consiste à alterner 
le calcul analytique des trajectoires de tous les grains jusqu'à occurrence du prochain choc, et le traitement d'une collision binaire 
avec modification des vitesses des deux grains concernés [39]. Le pas de temps n'est donc pas fixé. Cette méthode fonctionne bien dans le cas 
des milieux fortement agités. Des raffinements techniques permettent de limiter le coût des calculs pour un intervalle de temps donné à l'ordre  
$n\log n$ avec $n$ grains [40]. 

N'étant pas capable de traiter les contacts maintenus, une telle méthode est inadaptée 
(parce qu'elle se fonde sur un modèle physique inadéquat) lorsque le système s'approche d'un état d'équilibre ou 
dans les régimes d'écoulement dense. En effet, on trouve alors des amas de plus de deux grains en contact. 
En fait ces situations de « multicontact » maintenu sont fréquentes dès que la dissipation au sein du matériau 
devient importante [44]. Les collisions doivent alors être traitées en considérant l'ensemble du réseau des contacts 
car il s'agit de processus collectifs qui ne peuvent être réduits aux chocs binaires. 

\subsubsection{La dynamique des contacts}

La méthode de dynamique des contacts [45-47], qui manipule également des corps rigides, parvient à éviter ce problème en 
traitant de la même façon les contacts maintenus et les collisions multiples survenus pendant un pas de temps de durée fixée, 
moyennant une astucieuse formulation qui unifie loi de contact et loi de choc. La méthode manipule des percussions. 
Les lois de contact sont exprimées par deux graphes reliant ces percussions à des vitesses formelles, 
moyennes des vitesses juste avant et juste après le choc, pondérées par des coefficients qui modélisent l'inélasticité des chocs, 
pris égaux aux coefficients de restitution. Ce formalisme permet de décrire aussi bien des chocs binaires que des contacts multiples, 
et de décrire la modification du statut d'un contact (ouverture, fermeture, passage de glissant à non glissant ou réciproquement). 
Il s'agit d'une méthode générale, apte à décrire des états fortement agités comme des milieux denses, qui ignore délibérément les petites échelles, 
d'espace (la déformation d'un contact) et aussi de temps (le détail d'une séquence de collisions). Ses paramètres sont, outre l'inertie des grains, 
le coefficient de frottement et les deux coefficients de restitution. Elle peut permettre d'avoir des temps de calcul plus courts : 
le pas de temps est supérieur à celui des autres méthodes car un seul pas de temps peut prendre en charge une ou plusieurs collisions 
(même s'il s'agit des chocs successifs d'un même grain !). Sa mise en oeuvre numérique introduit toutefois subrepticement le pas de temps comme 
nouveau paramètre dans le modèle (puisqu'il décide du raffinement avec lequel on détaille une suite de collisions), ainsi qu'une petite imprécision 
du traitement de l'impénétrabilité. La détermination des percussions, via un algorithme implicite,  nécessite également un processus d'itération 
interne dont la durée tend à augmenter à l'approche de l'équilibre. Cette méthode a été appliquée avec succès à 
différents problèmes, quasi-statiques [48-51], ou dynamiques [44, 52, 53]. La méthode dite des percussions, introduite dans [20], 
en est assez proche. Une modification simple [54] permet de traiter les grains cohésifs : il suffit par exemple de rajouter, en plus 
des percussions lors des chocs, une force attractive, constante et égale à  $F_0$, entre toutes paires de grains dont les surfaces sont distantes 
de moins de $D_0$ . On a alors un modèle qui s'apparente plutôt au cas DMT (cf §2.1.4), puisque $h_0=0$  avec des grains rigides.

\subsubsection{Méthodes statiques}

D'autres approches, dédiées aux assemblages de grains avec un réseau de contacts établi a priori, ne traitent que de problèmes statiques, 
sans introduire l'inertie des grains ou toute autre forme de dynamique. Ces méthodes statiques [55, 56] font appel à la résolution de systèmes 
linéaires plutôt qu'à l'intégration d'équations différentielles, et s'apparentent au calcul aux éléments finis en élasticité ou en élastoplasticité. 
Elles déterminent la séquence des états d'équilibre atteints, sous un chargement lentement variable, par un réseau de contacts élastiques et 
frottants (ressorts et patins). Elles sont plus complexes à mettre en {\oe}uvre et moins générales que les méthodes dynamiques, 
puisqu'elles ne fonctionnent plus lorsque le réseau de contacts se réarrange brusquement, et ne peuvent traiter pour les 
assemblages granulaires usuels que le régime des très faibles déformations. Elles présentent cependant les avantages de comporter moins 
de paramètres et de permettre l'étude précise de la stabilité de configurations d'équilibre [55-57]. Elles sont naturellement adaptées à la 
simulation des assemblages de grains avec ponts solides comme les frittés [58], intermédiaires entre les assemblages granulaires et les 
matrices poreuses. 

\subsubsection{Méthodes de construction géométriques ou cinématiques}

Il s'agit simplement d'algorithmes de construction de configurations denses de grains non interpénétrés, 
censés reproduire une géométrie plausible d'empilement [59]. Les méthodes prenant en compte les propriétés 
mécaniques tendent actuellement à les supplanter.
 
\subsection{La démarche d'expérimentation numérique}

	Une fois choisi un modèle d'interaction entre grains et une méthode de simulation numérique, 
nous discutons maintenant les autres éléments qui font partie intégrante de la démarche d'expérimentation numérique. 
Il s'agit en premier lieu de bien définir le système étudié, les conditions aux limites qui lui sont appliquées, et son mode de préparation. 
Les remarques qui suivent portent sur la façon de traiter les résultats de l'expérience numérique, l'analyse statistique des données 
et l'analyse dimensionnelle qui permet de construire les paramètres pertinents.

\subsubsection{Conditions aux limites, sollicitations}

\begin{figure}[!htbp]
\bc
\includegraphics[height=9cm]{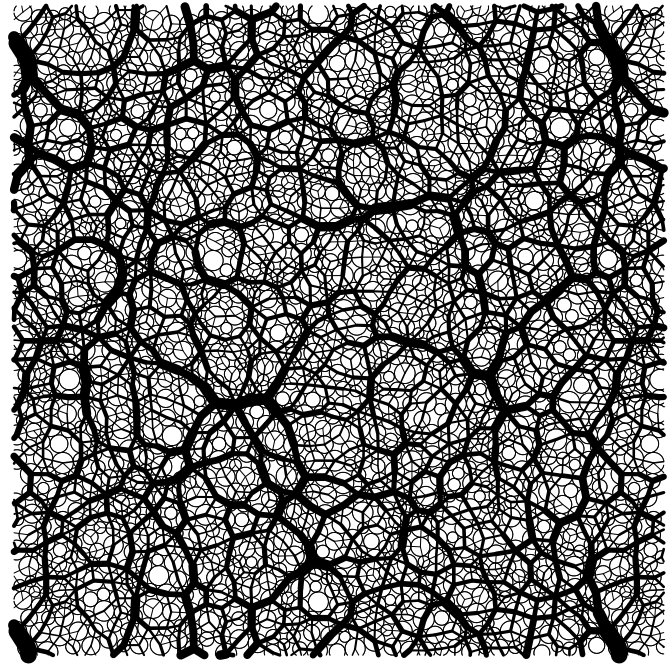}
\caption{Aspect des «~chaînes de force~» dans un échantillon bidimensionnel périodique de 1024 grains circulaires, 
en équilibre mécanique. L'épaisseur du trait joignant deux centres de disques en contact est proportionnelle à
l'intensité de la force normale.
\label{fig2}
}
\ec
\end{figure}

	Si l'on cherche à déterminer le comportement intrinsèque, en volume, du matériau, on souhaitera s'affranchir d'éventuelles parois, 
et l'on adoptera souvent des conditions aux limites périodiques [51, 60]. Les Fig.~\ref{fig2} et \ref{fig5}  
visualisent l'effet de conditions périodiques. 
On peut au contraire s'intéresser au comportement au voisinage d'une paroi, le matériau étant par ailleurs considéré comme uniforme dans 
la direction transverse, et l'on ne maintiendra alors le caractère périodique que dans certaines directions. On pourra ainsi étudier le 
comportement d'un matériau confiné dans une conduite, ou à surface libre sur un plan incliné [44]. 
On peut aussi choisir de décrire une géométrie 
plus compliquée telle qu'une cellule de cisaillement annulaire [61]. 
On devra par ailleurs indiquer les sollicitations imposées au matériau : 
lorsque l'on s'intéresse à sa rhéologie, on impose certaines composantes du tenseur des contraintes, 
et on mesure les taux de déformation ou les déformations associées, et/ou vice-versa. 
\subsubsection{Préparation du système}
A la différence des molécules qui, soumises à l'agitation thermique, adoptent au cours du temps différentes configurations spatiales, 
échantillonnées selon la statistique de Boltzmann [41], les grains macroscopiques tendent à évoluer vers des états d'équilibre 
(au sens purement mécanique plutôt que thermodynamique) qui dépendent de l'histoire des sollicitations appliquées. 
Pour la simulation comme pour l'expérience, on doit donc se préoccuper de l'influence du processus d'assemblage, dont on 
verra l'importance au §3.2 dans le cas quasi-statique. Dans le cas des écoulements, l'influence de l'état initial, 
déterminante dans la phase de démarrage, peut disparaître en régime stationnaire (cf. §4).

\subsubsection{Représentativité, limite des grands systèmes}

Comme pour d'autres matériaux désordonnés, l'identification d'un comportement mécanique macroscopique des matériaux 
granulaires passe par des expériences, qu'elles soient menées à bien en laboratoire ou sur ordinateur, 
avec des échantillons représentatifs. 
On doit donc se préoccuper des fluctuations de comportement observées entre échantillons différents, mais statistiquement similaires, 
obtenus en appliquant les mêmes sollicitations à des configurations initiales aléatoires différentes 
mais régies par les mêmes lois statistiques. 
Lorsque la taille des échantillons augmente, ces fluctuations devraient progressivement diminuer, 
et la réponse aux sollicitations doit converger vers une loi de comportement déterministe. Il est particulièrement important, 
en pratique, de vérifier cette approche de la limite des grands systèmes dans le cas des matériaux granulaires. 
En effet, tant la distribution spatiale des efforts (Fig. 2), avec les « chaînes de forces » remarquées pour la première fois par 
P. Dantu dans les années 50 [62] que celle des déplacements entre configurations voisines, ou des vitesses (Fig. 3) 
présentent des hétérogénéités à des échelles sensiblement supérieures à la taille des grains. L'application directe des méthodes de changement 
d'échelle dans les matériaux aléatoires [63] est ainsi mise en défaut. En écoulement, une fois obtenu un état stationnaire, 
on considère que les grandeurs d'intérêt (positions, vitesses, forces...) ont un comportement moyen indépendant du temps 
et on procède donc à une moyenne temporelle.
\begin{figure}[!htbp]
\bc
\includegraphics[height=8cm]{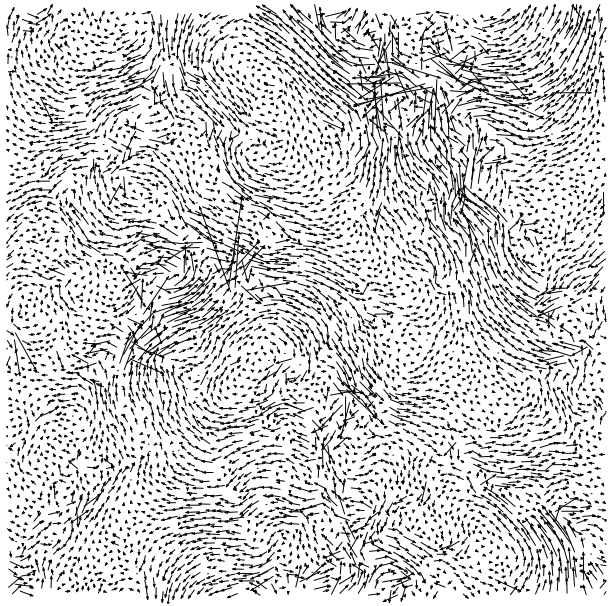}
\caption{Déplacements des centres des grains entre deux états d'équilibre voisins,
une fois soustraite la contribution de la déformation d'ensemble de l'échantillon.
\label{fig3}
}
\ec
\end{figure}

\subsubsection{Analyse dimensionnelle}

	Afin d'analyser les résultats, il faut commencer par faire la liste des paramètres décrivant le matériau et les sollicitations. 
On trouve ainsi pour le matériau : la taille d et la masse m des grains, la raideur des contacts $K_N$  (ou le module d'Young suivant que l'on 
prend des contacts élastiques linéaires ou hertziens), le coefficient de restitution $e_N$  (ou le paramètre $\zeta$, voir Eq. (10)), 
le coefficient de frottement $\mu$, et l'énergie interfaciale $\gamma$ des grains. (On choisira dans la suite $K_T = K_N /2$
pour les raideurs dans le cas linéaire, et  $\nu=0.3$ pour le coefficient de Poisson dans les contacts de Hertz,
afin de limiter le nombre des paramètres). 
Quant aux sollicitations, on trouve la pression $P$, le taux de déformation $\dot \epsilon$  ou  $\dot \gamma$, éventuellement la gravité $g$. 
On cherche alors à exprimer les résultats à partir des échelles naturelles de longueur ($d$), de masse ($m$) et de temps du système simulé. 
Plusieurs candidats existent pour l'échelle de temps : le temps de cisaillement ($1/\dot \gamma$),
le temps inertiel ($\sqrt{m/Pd}$ à 3 dimensions, $\sqrt{m/P}$ à 2 dimensions), 
le temps de collision ($\tau_c$) ou encore le temps de chute en présence de gravité ($\sqrt{d/g}$).
L'analyse dimensionnelle prédit alors que les comportements 
observés, si on les exprime par des grandeurs adimensionnées, dépendent des nombres sans dimension que l'on définit ci-dessous, 
et que l'on doit considérer comme des paramètres de contrôle. Pour mieux exploiter ce résultat mathématique, 
il est commode de définir ces nombres (qui sont des rapports d'échelles de longueur, de temps ou de force) de telle sorte 
que l'on pourra trouver des quantités possédant le même sens physique (à un facteur numérique d'ordre 1 près) 
pour des modèles de contacts différents. 

Le niveau de raideur des grains $\kappa$ est inversement proportionnel à la déflexion normale typique dans les contacts, divisée par le diamètre, 
dans un échantillon à l'équilibre sous la pression $P$. 
Pour la loi de Hertz, une définition en est $\kappa=(E/P)^{2/3}$ [56] et pour l'élasticité linéaire (2D)  $\kappa=K_N/P$.  $\kappa\to \infty$
correspond à la limite des grains rigides. 

Le nombre d'inertie $I=\dot \gamma \sqrt{m/Pd}$ ou $\dot \epsilon \sqrt{m/Pd}$ 
quantifie l'importance des effets dynamiques. Il mesure le rapport des forces d'inertie et des forces imposées : 
une petite valeur correspond au régime quasi-statique, tandis qu'une grande valeur correspond au régime inertiel ou encore « dynamique ». 

L'intensité de la cohésion, relativement au niveau de pression imposée, peut être exprimée par $h_0\kappa/d$, rapport de $h_0$
(défini au §2.1.4) sur la déflexion élastique du contact due à la pression, (c'est aussi $ F_0/Pd^2$ en utilisant la force adhésive). 
Pour le modèle 2D introduit à la fin du §2.1.4 (Eq. (9)), on peut introduire le nombre sans dimension $\eta=\gamma^2 /(K_NP)$. 
Considérant le rapport entre la résistance à la traction et la force moyenne appliquée, la transition entre les régimes de 
faible et de forte cohésion doit dépendre du produit $\eta$.
(En général, un modèle de cohésion introduit un paramètre sans dimension supplémentaire, 
le rapport défini en (8) - ou bien le paramètre de Tabor (7) - qui est infini dans ce modèle simple).

\section{Assemblages solides, sols granulaires}
\begin{figure}[!htbp]
\bc
\includegraphics[height=7cm]{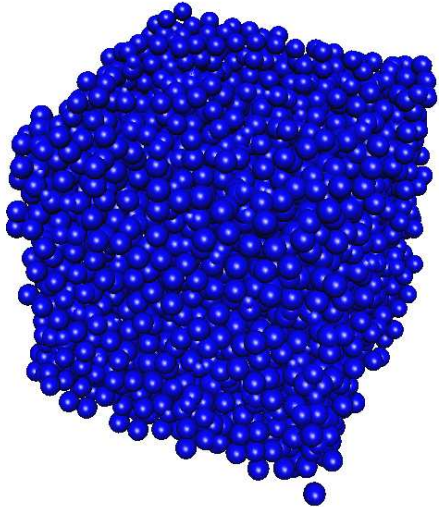}
\caption{Vue en perspective d'un échantillon cubique de 4000 billes.
\label{fig4}
}
\ec
\end{figure}
La simulation numérique discrète est fréquemment utilisée pour la compréhension des comportements mécaniques de matériaux pulvérulents 
tels qu'étudiés en mécanique des sols [6].  Nous illustrons ici ses apports dans le cas de la compression triaxiale d'une collection de billes 
de verre sphériques, que nous supposerons toutes de même diamètre d (Fig. 4). La Fig. 5 en montre une vue en coupe. 
Un tel matériau modèle se prête aux mêmes mesures en laboratoire que les sables. Après avoir introduit la problématique (§3.1), 
nous présentons des résultats relatifs à la préparation des échantillons (§3.2),
étudions le rôle des différents paramètres micromécaniques (§3.3), 
discutons de la nature des déformations et de la définition de modules élastiques (§3.4),
ainsi que de la caractérisation de l'état 
interne du matériau (§3.5). Le §3.6 est une brève conclusion.
\begin{figure}[!htbp]
\bc
\includegraphics[height=7cm]{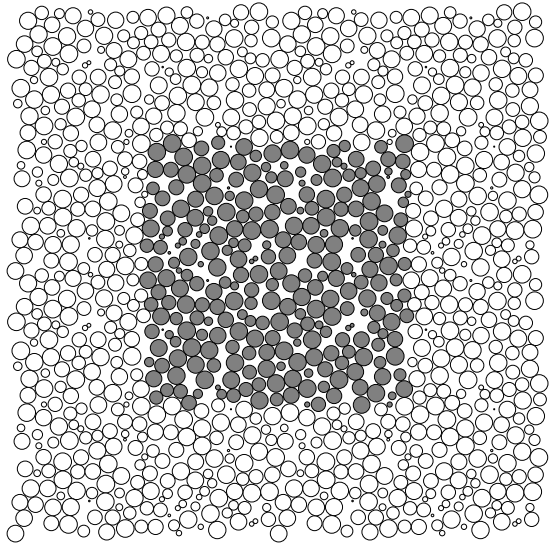}
\caption{Vue en coupe d'un échantillon tridimensionnel périodique de billes,
le plan de coupe étant parallèle à l'un des plans de symétrie de la cellule cubique.
Les grains de la cellule de base sont figurés en gris, les autres en sont des images par translation.
\label{fig5}
}
\ec
\end{figure}

\subsection{Paramètres du matériau modèle, premier exemple d'expérience numérique}
Nous attribuons ici, sauf indication contraire, un coefficient de frottement $\mu=0.3$
aux contacts entre particules, dont le comportement 
élastique est conforme aux équations (1-3) avec $E=70$GPa et $\nu=0.3$, 
valeurs appropriées pour des billes de verre. Les calculs numériques sont conduits par 
la méthode de dynamique moléculaire décrite au §2.2.1. Pour certaines configurations en équilibre mécanique,
on a également, par une approche 
statique au sens du §2.2.4 ci-dessus (construction de la matrice de raideur associée au réseau des contacts)
déterminé les modules élastiques tangents dans la limite des très faibles déformations.
\begin{figure}[!htbp]
\bc
\includegraphics[height=8cm]{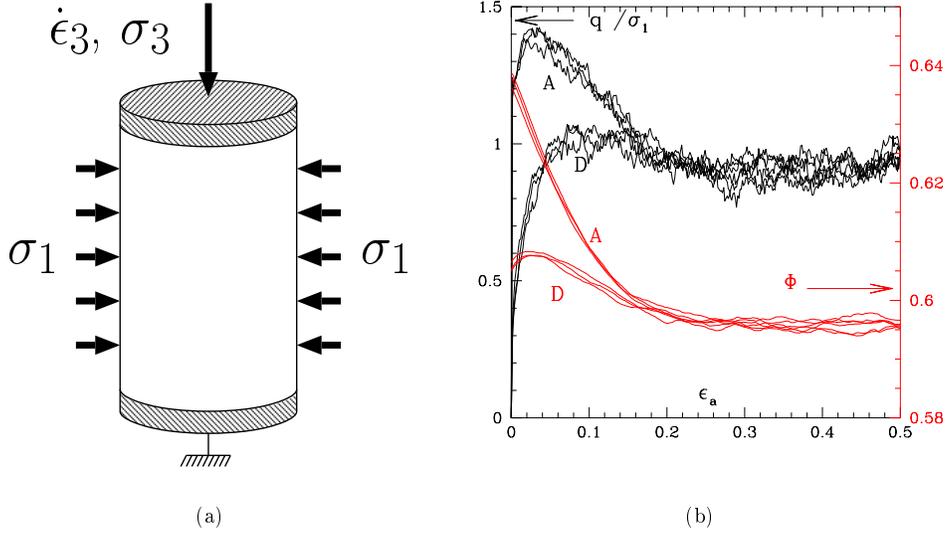}
\caption{Essai triaxial.
(a) Schéma du test de compression triaxiale. La  contrainte latérale $\sigma_1=\sigma_2$ 
est maintenue égale à la pression de confinement $P$. 
(b) Courbes $q/P$ (axe à gauche) et $\Phi$ (axe à droite) fonctions de la déformation axiale  $\epsilon_a = \epsilon_3$
obtenues par simulation numérique de 3 échantillons de 4000 billes de type A,
préparés dans un état initial isotrope très dense, et 3 échantillons de type D, plus lâches. 
A noter l'approche du même état critique. Ici $P=100$kPa et $\mu=0.3$ .
\label{fig6}
}
\ec
\end{figure}

Le test de compression triaxiale, classique en mécanique des sols [8], est schématisé sur la Fig. 6a.
Il consiste à placer une éprouvette cylindrique entourée d'une membrane souple, qui transmet la pression $P$
d'un fluide de confinement, entre les deux plateaux d'une presse qui 
imposent une contrainte verticale   différente de P, les contraintes horizontales $\sigma_1=\sigma_2$ 
restant égales à $P$. Le dispositif est axisymétrique, et les directions principales de contrainte
et de déformation sont constantes (horizontales et verticales sur le schéma). On a coutume de mesurer, 
en fonction de la déformation axiale   de l'échantillon (raccourcissement positif), le déviateur des contraintes
$q=\sigma_3-\sigma_1$ et la déformation volumique $ \epsilon_v = 1 - (1-\epsilon_1)(1-\epsilon_2)(1-\epsilon_3)$. 

Une simulation numérique de compression triaxiale peut avoir deux objectifs :
soit chercher à imiter l'expérience de laboratoire, ce qui permet éventuellement d'en quantifier
les imperfections, qui conduiront à des inhomogénéités (frettage des parties haute et basse, 
rôle de la membrane...) ; soit mesurer autant que possible une loi de comportement, en mettant en {\oe}uvre un test « idéal ». 
C'est cette deuxième démarche qui est adoptée dans les études présentées ici.
Afin de supprimer les effets de bord et de s'approcher plus rapidement 
de la limite macroscopique, on a adopté des conditions aux limites périodiques.
Les résultats ci-dessous sont obtenus sur des échantillons de 4000 particules, tels que celui de la Fig. 4. 

La Fig. 6b montre l'évolution caractéristique du déviateur et de la déformation volumique avec $\epsilon_a$
dans le cas d'un état initial dense (A) et d'un autre plus lâche (D). Ces résultats, obtenus par simulation numérique,
montrent que l'on retrouve bien les comportements classiquement observés en laboratoire. Le déviateur,
lors de la compression triaxiale de l'échantillon dense, passe par une valeur maximale $q_{pic}$
(« pic » de déviateur), associé à une valeur d'angle de frottement interne  $\varphi_{pic}$ par
$$
\frac{q_{pic}}{P}=\frac{2\sin\varphi_{pic}}{1-\sin\varphi_{pic}}.
$$
Il décroît ensuite pour atteindre un plateau, 
tandis que la déformation volumique, très rapidement négative
(le matériau se dilate après une très courte phase initiale contractante invisible 
sur la Fig. 6b) sature à une valeur asymptotique. Pour les échantillons D, plus lâches, le déviateur $q$ augmente graduellement vers 
la même valeur asymptotique que dans le cas A, tandis que la dilatance, moins forte (des échantillons encore plus lâches 
resteraient contractants) est telle que la compacité (fraction volumique occupée par le solide)  $\Phi$ tende vers la même limite   
 $\Phi_c$ que dans le cas A. On retrouve donc par la simulation la notion d'état critique, état limite rejoint par le matériau, 
indépendamment de son état initial, pour une sollicitation monotone produisant des déformations assez grandes, 
et caractérisé par la poursuite de la déformation plastique sans changement de volume [8]. 

Ce premier exemple indique que la simulation numérique permet d'observer les mêmes comportements que les expériences de laboratoire. 
On a ici un angle de frottement interne $\varphi_c$ d'environ 20 degrés, et une compacité critique $\Phi_c$ proche de 0.596 
(l'indice des vides critique est bien sûr $n_c = \frac{1-\Phi_c}{\Phi_c}$ ). On trouvera par exemple dans [60] une étude numérique, 
dans un système similaire, de l'influence de $\mu$ sur $\Phi_c$ .

Les résultats de la Fig. 6b sont encourageants puisqu'ils reproduisent des comportements expérimentalement connus. 
Au-delà de ce constat rassurant, il faut s'interroger cependant sur leur sensibilité à la procédure d'expérimentation 
numérique et au choix des paramètres, et aussi les exploiter pour comprendre les origines microscopiques des phénomènes 
macroscopiques observés. Notons toutefois que la taille des échantillons et les conditions aux limites périodiques choisies empêchent 
la localisation de la déformation.

\subsection{Préparation d'échantillons solides : compacité et nombre de coordination}

On a coutume de caractériser expérimentalement l'état initial d'un sable par sa compacité (ou son indice des vides),
grandeur accessible à la mesure, et qui influe fortement sur le comportement, comme on vient de le rappeler.
Les échantillons de sable sont préparés de façon contrôlée par différentes techniques, comme la pluviation [64] ou le damage humide. 

Numériquement, on doit également commencer par fabriquer un assemblage de particules en contact, susceptible de supporter des contraintes. 
Là aussi, deux voies sont possibles.
On peut tenter de  reproduire numériquement la procédure expérimentale. Cela donne des calculs assez complexes, 
car les méthodes de laboratoire font intervenir une multitude de phénomènes (rôle de la gravité, rôle des parois) dans des échantillons 
dont l'homogénéité sera nécessairement imparfaite. Une étude numérique de la pluviation est en cours [65]. Les travaux numériques menés à bien 
jusqu'ici [51, 60], et ceux dont les résultats sont présentés ci-dessous, ont plutôt suivi une autre approche : les méthodes de préparation 
sont spécifiquement adaptées à la simulation numérique, elles restent similaires aux procédures du laboratoire, mais leur mise en {\oe}uvre
expérimentale directe serait impraticable. 

L'étude des échantillons ainsi préparés permet d'étudier et de classifier les états possibles des matériaux granulaires solides. 
Ainsi assemble-t-on des particules en partant d'un état initial de compacité faible (« gaz granulaire »), 
sans aucun contact, par \emph{compactage homogène et isotrope} d'un échantillon périodique [60]. Soit la cellule périodique dans laquelle 
est enclos l'échantillon se réduit, soit les particules gonflent dans un récipient fixe. Lors de cette phase d'assemblage, on peut 
éventuellement adopter une valeur $\mu_0 $ plus faible du coefficient de frottement :
l'état produit sera \emph{a fortiori} capable de supporter les efforts appliqués avec la valeur finale de $\mu$.
On constate alors que la compacité de l'état d'équilibre final, sous pression 
isotrope P fixée, est d'autant plus élevée que $\mu_0$ est faible. En prenant, cas extrême, $\mu_0=0$, 
on obtient pour les sphères la compacité maximale des assemblages aléatoires
(\emph{random close packing}[59, 66]), dont la valeur $\Phi_{RCP}\simeq 0.638$ est bien connue. Ce procédé nous donne les assemblages 
de type A de la Fig. 6b, avec un nombre de coordination  $z^*$ (nombre moyen de contact par grain) de 6 pour les seuls grains actifs 
(en l'absence de force imposée autre que la pression extérieure, une certaine fraction des grains, faible dans ce cas, mais pouvant 
atteindre 10 ou 15\% dans les échantillons moins bien coordonnés, ne participe pas à la transmission des efforts. 
Ce phénomène est visible à deux dimensions sur la Fig. 2). A l'opposé, si on choisit $\mu_0=\mu$, 
on obtient l'état plus lâche D de la Fig. 6b, pour lequel $\Phi=0.606$ et $z^*\simeq 4.62$ .
Ces valeurs correspondent ici à la pression $P = 10$kPa, soit  $\kappa \simeq 37000$ pour la rigidité
adimensionnelle introduite au §2.3.4. On peut concevoir que le choix d'un coefficient de frottement très faible  $\mu_0=0.02$
lors de l'assemblage est une modélisation simple d'un compactage 
en présence d'un lubrifiant dans les contacts. L'état obtenu sera appelé B ci-dessous.

On peut enfin chercher à reproduire, sous forme simplifiée, le compactage par vibrations, dans lequel l'ouverture intermittente 
des contacts limite la mobilisation du frottement. A cette fin, on a dilaté, par une homothétie sur les positions de rapport $1.005$, 
les configurations A, supprimant ainsi tous les contacts, puis on a mélangé à volume constant, pour compacter finalement jusqu'à l'équilibre 
sous la pression de consigne $10$kPa avec $\mu=0.3$ . On obtient ce faisant l'état C. Il est remarquable que cet état est plus dense que B, 
avec une compacité de $0.635$, proche de $\Phi_{RCP}$, tandis que le nombre de coordination est proche de celui de l'état
le plus lâche, D :  $z^*\simeq 4.56$ contre  $z^*\simeq 5.80$ pour B.
On voit donc que compacité et nombre de coordination peuvent, pour ces états isotropes, varier indépendamment.
\begin{figure}[!htbp]
\bc
\includegraphics[height=10cm]{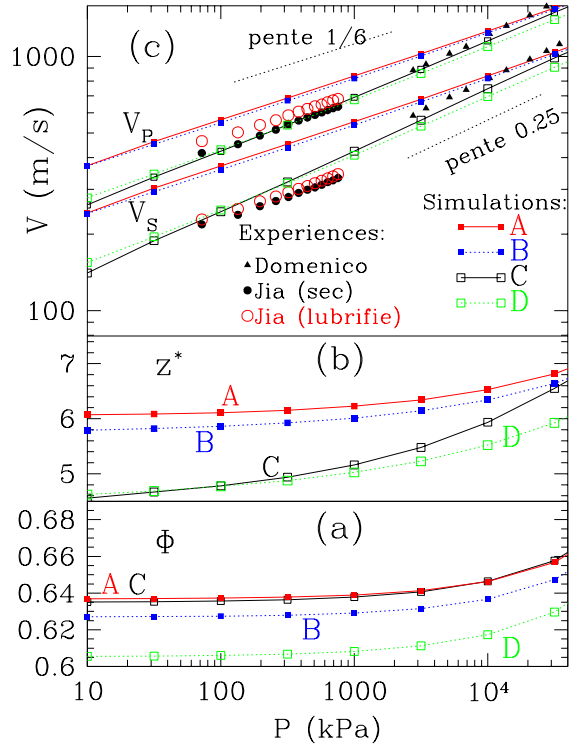}
\caption{Compacité  $\Phi$ (a), nombre de coordination $z^*$  (b) et vitesses des ondes longitudinales
(P) et transversales (S) dans les échantillons numériques préparés dans les états A, B, C et D.  $V_P$ et $V_S$ 
sont comparées aux valeurs mesurées par Domenico [67] et par Jia et Mills [68]
avec des billes de verre sèches (préparation par vibration) et lubrifiées.
\label{fig7}
}
\ec
\end{figure}
La simulation de la compression isotrope au-delà de la pression initiale $10$kPa, avec $\mu=0.3$, et jusqu'à $100$MPa (en supposant, 
ce qui est peu probable en pratique pour les pressions les plus élevées, que les contacts restent élastiques) révèle une 
augmentation quasi-réversible de  $\Phi$ et de $z^*$ avec $P$ (Fig. 7). Le calcul des modules élastiques de compression $B$
et de cisaillement $G$ permet de comparer les vitesses des ondes élastiques $V_P=\sqrt{(B+\frac{4}{3}G)/\rho}$ et
$V_S=\sqrt{G/\rho}$ ($\rho$ est la masse volumique du matériau granulaire) à des mesures sur 
échantillons de billes de verre [67, 68], assemblées soit par tassement avec secousses répétées, soit en présence d'une faible quantité 
d'un lubrifiant solide.
Les résultats de la Fig. 7 montrent que ces vitesses sont sensibles au nombre de coordination plutôt qu'à la compacité, 
et que les échantillons C se rapprochent, pour la vitesse des ondes et sa variation avec P,
des échantillons expérimentaux vibrés, tandis que les configurations B ont un comportement effectivement
similaire à celui des billes lubrifiées (dont l'assemblage en laboratoire est moins compact, 
mais avec une vitesse des ondes plus élevée). 
\begin{figure}[!htbp]
\bc
\includegraphics[width=14cm]{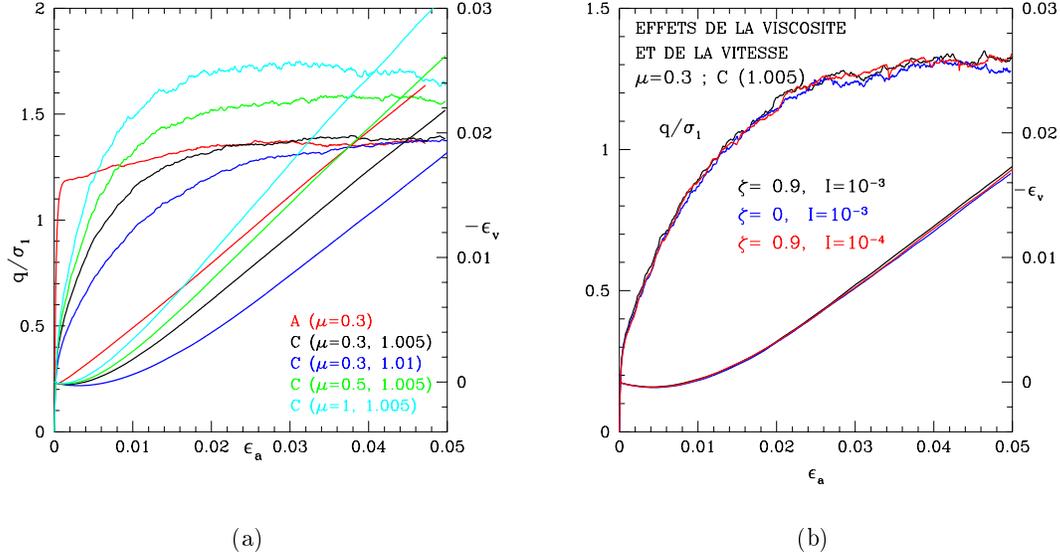}
\caption{
(a) Effet de l'assemblage initial (A ou C, nombre de coordination différent)
et du coefficient de frottement intergranulaire $\mu$ ($0.3$, $0.5$, ou $1$) sur la réponse des échantillons au test triaxial ($P=100$kPa).
Pour chaque état initial et valeur de  $\mu$ les courbes sont des moyennes sur 5 échantillons.
(b) Mise en évidence de l'absence d'influence des paramètres dynamiques $I$ et $\zeta$ dans la limite quasi-statique.
\label{fig8}
}
\ec
\end{figure}
Cette variation de nombre de coordination qui ne suit pas celle de la compacité, et qui semble donc pouvoir être décelée par des 
mesures de modules élastiques en très faibles déformations, entraîne une réponse mécanique différente, pour les faibles déformations, 
dans une compression triaxiale (Fig. 8a). Même si $q_{pic}$ et $\varphi_{pic}$ sont les mêmes, aux incertitudes près, 
pour les états initiaux A ($z^*\simeq 6$) et C ( $z^*\simeq 4.5$), la montée du déviateur, dans le cas A,
et le passage de la contractance à la dilatance sont beaucoup plus rapides. Les configurations de type C 
demandent une déformation axiale de l'ordre de $2$ ou $3\%$ pour mobiliser le frottement interne tg$\varphi _{pic}$.
Les courbes mesurées en laboratoire avec les sables ou les billes sont beaucoup plus proches,
en général, du cas C que du cas A. Sous réserve d'études complémentaires de 
l'anisotropie initiale qui affecte les échantillons expérimentaux [15, 65], la confrontation expérimentale confirme donc que les états 
initiaux denses de grains secs sont plutôt de type C, avec un faible nombre de coordination.

\subsection{Compression triaxiale d'échantillons denses : brève étude paramétrique}

Le coefficient de frottement intergranulaire $\mu$ influe sensiblement sur la valeur « au pic »
de l'angle de frottement interne $\phi_{pic}$  (Fig. 8a). 
Par ailleurs, l'analyse dimensionnelle prédit que $q/P$  et $\epsilon_v$ , fonctions de $\epsilon_a$
  dans la compression triaxiale monotone, ne devraient dépendre que 
des paramètres sans dimension $I$, $\kappa$, $\mu$ et $\zeta$.
De plus, dans la limite quasi-statique ($I$ petit), l'influence de $I$ et de $\zeta$ devrait disparaître. Enfin, on 
peut se demander si $\kappa$ reste pertinent lorsqu'il est très grand (limite des grains rigides). Les Fig. 8b, 9a et 9b montrent effectivement 
qu'à l'échelle des déformations de l'ordre de 1/100 ou 1/1000, seul $\mu$ influe sur le comportement, pourvu que l'on ait  $\kappa> 1700$
(valeur pour $P=1$MPa ) et $I\le 10^{-3}$  (dans le cas considéré d'échantillons denses).
L'absence de sensibilité au paramètre d'inertie $I$ (limite quasi-statique) 
laisse supposer que l'on peut simuler des expériences de laboratoire beaucoup plus lentes ($I$ prenant des valeurs de $10^{-8}$ ou
$10^{-9}$ avec les taux de déformation usuels de $10^{-4}$ ou $10^{-5}\mbox{s}^{-1}$). 
Comme il sied à un angle de frottement,  $\phi_{pic}$ est indépendant de la pression de confinement dans la 
gamme de raideurs $\kappa$ étudiée, et il en est de même pour la dilatance  $-\frac{d\epsilon_v}{d\epsilon_a}$ .
\begin{figure}[!htbp]
\bc
\includegraphics[height=8cm]{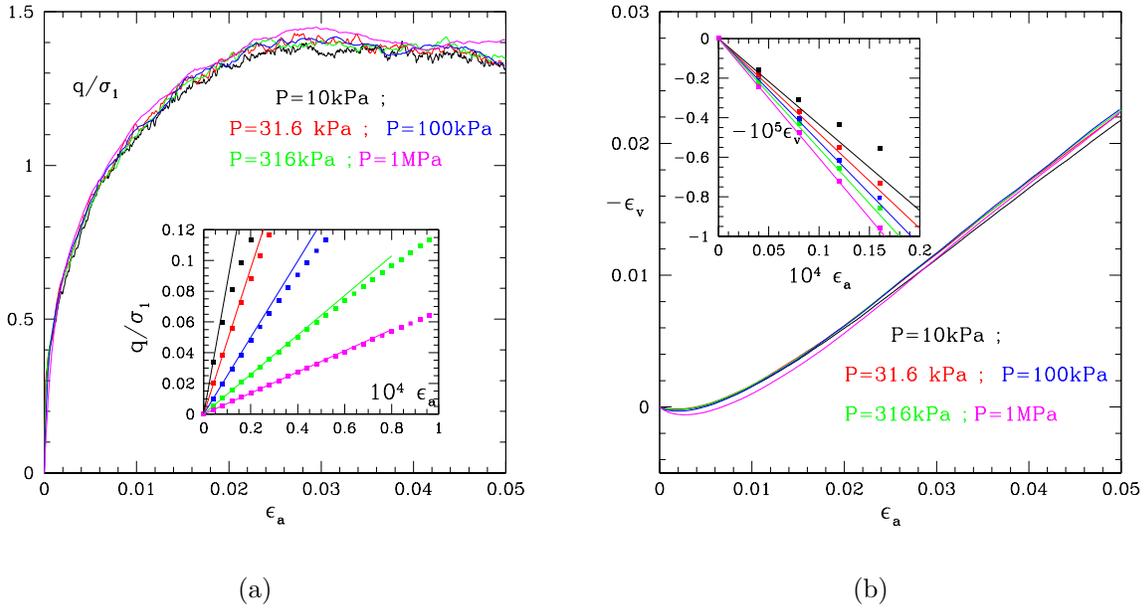}
\caption{Influence de la pression (ou du paramètre sans dimension $\kappa$) sur les courbes $q/P$ (a)  et $\epsilon_v$ (b) fonctions de
$\epsilon_a$ (échantillon de type C, dilatation 1.005). Les petits schémas insérés sont des agrandissements du voisinage immédiat de 
l'origine en déformation (état initial isotrope) et les pentes des droites sont données par le calcul 
des modules élastiques par une méthode statique (assemblage de la matrice de raideur), 
que l'on trouve effectivement tangentes aux courbes obtenues lors de la simulation du test triaxial 
(résultats figurés par les points). $\kappa$, sans influence sur $q/P$ et  $\epsilon_v$ pour $\epsilon \sim 0.01$, 
affecte la réponse mécanique pour $\epsilon \sim 10^{-5}$.
\label{fig9}
}
\ec
\end{figure}
Revenons enfin sur la représentativité des échantillons. La Fig. 6b donne une indication des différences entre 
échantillons préparés dans le même état macroscopique, avec $n=4000$ billes (noter les plus faibles fluctuations avant le pic). 
En revanche, avec $n=1372$ grains seulement, les fluctuations d'un échantillon
à l'autre illustrées par la Fig. 10 sont considérablement plus élevées. 
En dimension 2, on a pu observer [57] que les écarts au comportement moyen diminuaient comme $n^{-1/2}$  pour $n\to +\infty$.
\begin{figure}[!htbp]
\bc
\includegraphics[height=8cm]{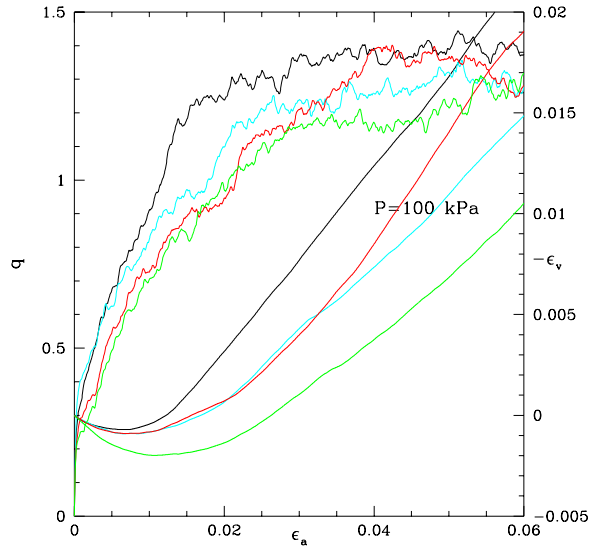}
\caption{Résultats de tests triaxiaux simulés sur 4 échantillons différents de 1372 billes
préparés dans le même état~: défaut de reproductibilité.
\label{fig10}
}
\ec
\end{figure}

\subsection{Nature de la déformation : quand mesure-t-on des modules élastiques ?}
Les modules élastiques correspondant aux vitesses des ondes représentées sur la Fig. 7 diffèrent manifestement de la pente des courbes 
contraintes/déformations de la Fig. 8 (qui correspondent au régime  $\epsilon \sim 0.01$
souvent modélisé comme élasto-plastique avec écrouissage). 
Leur valeur est bien supérieure (plusieurs centaines de MPa pour $P = 100$kPa) et ils varient avec la pression $P$ approximativement comme 
une loi de puissance avec un exposant entre $1/3$ et $1/2$ (Fig. 7). En revanche, $\sigma_3$, 
pour des déformations modérées, est proportionnel à $P$
dans la compression triaxiale monotone. Il faut se placer dans la gamme des très faibles déformations, comme illustré dans les graphes en 
encart de la Fig 9, pour que la pente $\frac{dq}{d\epsilon_a}$ se rapproche du module d'Young $E$,
et la pente $\frac{d\epsilon_v}{d\epsilon_a}$ de $1-2\nu$.  Ces résultats sont conformes aux observations 
expérimentales [69]. Lorsque l'on adopte un modèle simple d'élastoplasticité avec critère de Mohr-Coulomb sans écrouissage, le « module » 
que l'on utilise n'est pas physiquement un vrai module élastique. C'est l'origine microscopique de la déformation macroscopique qui explique 
les différences entre ces « modules » apparents. La déformation du matériau est essentiellement due à la déformation des contacts 
dans le régime des très petites déformations (où l'on peut observer le module élastique), tandis qu'elle résulte de ruptures du réseau 
des contacts, immédiatement suivies de réparations, pour les déformations plus élevées. Pour autant, ne mesure-t-on des modules élastiques 
qu'au voisinage de l'état initial isotrope ? C'est ce qui semblerait ressortir de l'aspect de la courbe du déviateur fonction de la 
contrainte axiale, qui ne possède de pente assez élevée qu'à l'origine. En fait, on peut aussi mesurer de « vrais » modules au 
voisinage d'autres points de la courbe, mais soit en décharge, soit en laissant au système un temps de maturation à contrainte constante, 
ou encore en le soumettant à des cycles répétés de très faible amplitude. Ces dernières procédures entraînent un certain fluage [69] 
préalable à la mesure de modules « statiques », en charge comme en décharge, dont les valeurs coïncident avec celles des modules 
« dynamiques » (déduits de la vitesse des ondes). 

La Fig. 11 montre l'effet de décharges à partir de différents états de compression triaxiale, ainsi que les modules 
élastiques correspondants, qui portent alors la trace de l'anisotropie induite par la déformation. 
\begin{figure}[!htbp]
\bc
\includegraphics[height=8cm]{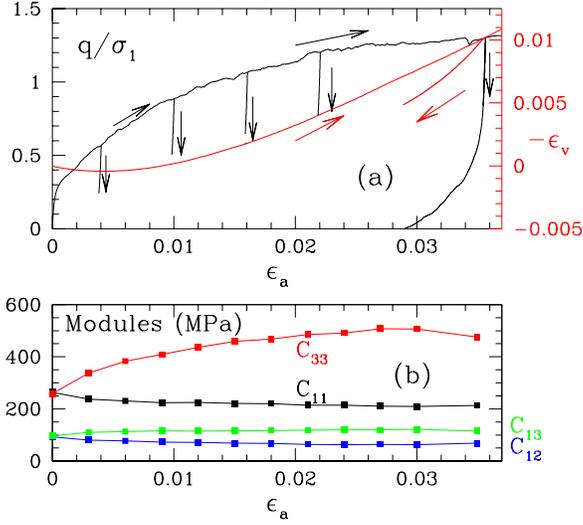}
\caption{(a) Effet de décharges à partir de différents états intermédiaires dans la compression biaxiale d'un échantillon de type C. 
On peut remarquer la grande raideur des courbes, qui ne s'écartent que lentement de leurs tangentes ayant pour pentes
les modules élastiques en très petites déformations.
(b) Modules élastiques (notation de Voigt) à différentes étapes de la compression triaxiale monotone, fonction de la déformation axiale. 
Mise en évidence de l'anisotropie induite.
\label{fig11}
}
\ec
\end{figure}
La simulation numérique permet donc de distinguer les déformations macroscopiques issues de la déformation des 
contacts de celles qui proviennent de réarrangements de la structure granulaire. Le rôle des caractéristiques microscopiques diffère 
suivant que l'une ou l'autre des origines physiques de la déformation est dominante. On trouvera dans [56, 70],  
des études plus détaillées des conséquences physiques (dont le phénomène de fluage) de ces deux régimes de comportement.

\subsection{Anisotropie et texture}
L'anisotropie induite par la déformation est apparente dans la distribution des orientations de contact. Au fur et à mesure 
que l'échantillon est comprimé dans la direction 3, la proportion de contacts parallèles à cette direction de compression 
s'accroît, tandis que les contacts orientés perpendiculairement se raréfient. Cette distribution, dans l'expérience axisymétrique que 
nous discutons ici, s'exprime par la densité de probabilité $P(\cos\theta)$ (fonction paire de $\cos\theta$, 
constante dans le cas isotrope), $\theta$  désignant l'angle entre le vecteur unitaire normal ${\bf n}$ au contact
et l'axe de compression (indice 3). On peut obtenir une approximation de $P(\cos\theta)$  
avec les trois premiers termes du développement en polynômes de Legendre dont les coefficients
sont donnés par les moments $\stav{n_3^2}$ et $\stav{n_3^4}$ . 
La Fig. 12a présente l'évolution de $P(\cos\theta)$  lors de la compressions triaxiale
(partie « pré-pic » dans le cas de la préparation type C au sens du §3.2). 
Les résultats de la Fig. 11 indiquent que l'on peut éventuellement retrouver cette anisotropie grâce à des mesures de modules élastiques. 
La Fig. 12b montre l'évolution de $z^*$, $\stav{n_3^2}$ et $\stav{n_3^4}$ vers les valeurs qui caractérisent l'état critique,
pour les échantillons de la Fig. 6b. 
L'état critique est en effet caractérisé par des valeurs de la compacité et du nombre de coordination,
mais aussi par la forme de $P(\cos\theta)$, 
résumée par les deux coefficients $\stav{n_3^2}$ et $\stav{n_3^4}$ , qui est associée à la « structure d'écoulement » plastique.
\begin{figure}[!htbp]
\bc
\includegraphics[height=8cm]{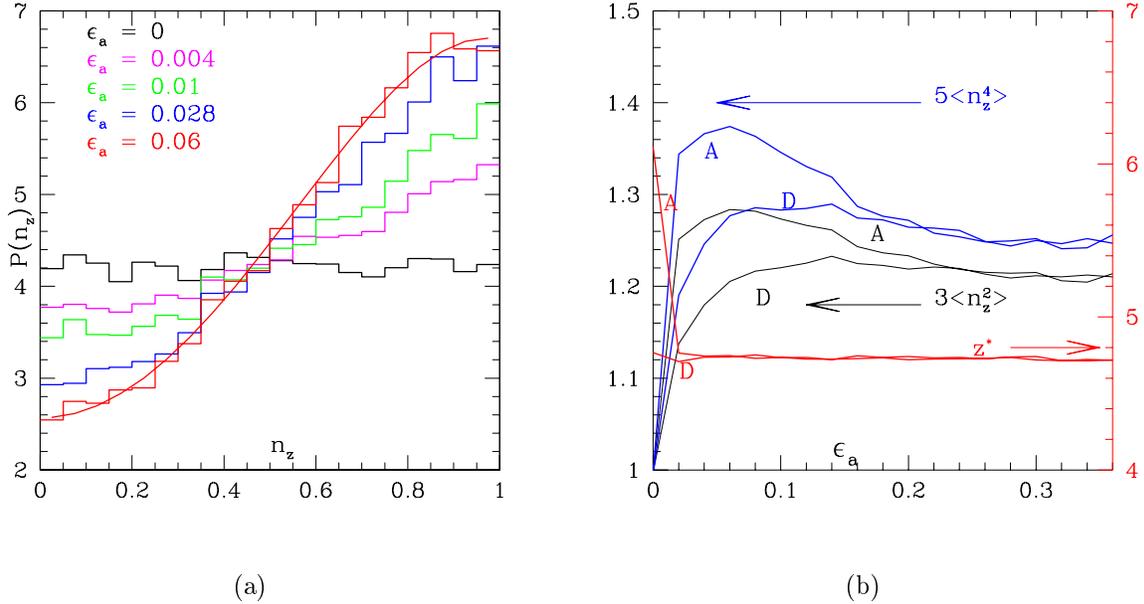}
\caption{(a) Distribution des orientations de contacts (représentée par un histogramme dont la somme donne le nombre de coordination)
à différentes étapes de la compression triaxiale d'un échantillon de type C (1.005). 
La courbe continue rouge figure pour $\epsilon_a=0.06$  la distribution paramétrée par $\stav{n_3^2}$ et $\stav{n_3^4}$.
(b)  Evolution des paramètres $z^*$, $\stav{n_3^2}$ et $\stav{n_3^4}$ à l'approche de l'état critique pour les simulations de la Fig. 6b.
\label{fig12}
}
\ec
\end{figure}
De nombreuses études numériques en dimension 2 [51] ont porté sur cette distribution angulaire (encore appelée \emph{texture}), 
variable interne candidate à une description du comportement du matériau intégrant une part d'information micromécanique. 
En dimension 3, on peut aussi espérer qu'un paramétrage plus général de $P(\theta,\varphi)$  (non nécessairement axisymétrique), ainsi qu'une 
modélisation de son évolution sous l'effet de déformations accumulées, débouche sur la formulation de lois constitutives 
prenant en compte la microstructure (par exemple via la définition de variables d'écrouissage).
\subsection{Conclusion}
La simulation numérique discrète donne accès à l'état interne des matériaux granulaires caractérisés par des variables, 
indépendantes de l'indice des vides, qui ne sont pas directement accessibles à l'expérience en laboratoire : nombre de coordination, 
texture. Les résultats numériques suggèrent néanmoins que des informations utiles pourraient être déduites des mesures de module élastique,
soit par le contrôle de très faibles déformations, soit par propagation d'ondes.  Enfin, la distinction de deux régimes de comportement, 
qui diffèrent par l'origine de la déformation macroscopique, est une voie vers la compréhension de l'accumulation progressive de 
déformations sous sollicitation cyclique. 

\section{Ecoulements denses}

Cette partie est consacrée aux écoulements granulaires, dont la compréhension constitue un enjeu important tant en géophysique 
(propagation d'avalanches, mouvement de dunes, glissement des failles) que dans l'industrie (manutention, mise en {\oe}uvrede poudres, 
granulats en génie civil, génie chimique, agroalimentaire, pharmacie, tribologie...). Nous ne discuterons que des écoulements denses, 
régime « liquide » intermédiaire entre le régime « solide » des déformations quasi-statiques (décrit au §3) et le régime « gazeux » des 
milieux agités et dilués [10]. Dans ce régime, l'écoulement est proche de se bloquer [71]; son comportement présente donc des analogies 
avec celui d'autres fluides à seuil, constitués de particules plongées dans un fluide interstitiel [72]. La loi de comportement et les mécanismes 
de blocage ne sont pas encore bien compris. Cependant des avancées sont venues durant la dernière décennie de la combinaison de simulations 
numériques discrètes et d'expériences sur des matériaux modèles dans des géométries simples [73, 74]. 

Nous évoquons ici les apports de la simulation numérique discrète dans deux géométries. Après avoir décrit les systèmes étudiés (§4.1), 
nous présentons l'étude d'un écoulement confiné en cisaillement plan (à pression et vitesse imposées), qui nous permet de mesurer la loi de 
comportement pour des grains secs puis cohésifs (§4.2). Nous voyons ensuite (§4.3)comment cette loi doit être complétée dans le cas d'un 
écoulement à surface libre sur plan incliné  (contrainte imposée). Cette description peut essentiellement se résumer dans une loi de frottement, 
exprimant la dépendance du coefficient de frottement effectif du matériau en fonction de son état de cisaillement. Le §4.4 est une brève conclusion.

\subsection{Description des expériences numériques}

Le matériau granulaire est un matériau modèle bidimensionnel, constitué d'une assemblée de $n$ (entre 1000 et 5000) disques légèrement 
polydisperses de diamètre moyen d ($\pm 20\%$) et de masse moyenne $m$. Ces grains sont caractérisés par un coefficient de frottement 
microscopique $\mu$ (égal en général à $0.4$), un coefficient de restitution $e_N$ et une énergie interfaciale $\gamma$. 

Deux méthodes de simulation numérique discrète ont été utilisées : la dynamique moléculaire (cisaillement plan, arrêt de 
l'écoulement sur plan incliné) [61] et la dynamique des contacts (écoulement stationnaire sur plan incliné) [44]. En dynamique 
moléculaire, on s'est placé dans la limite des grains rigides ($\kappa \simeq 10^4$), et on a choisi en général $e_N = 0.1$.
En dynamique des contacts, $e_N = e_T = 0$. On a vérifié dans quelques cas le bon accord des deux méthodes
en comparant des systèmes identiques simulés par l'une ou l'autre méthode.

Contrairement au §3, le matériau est confiné par une (plan incliné) ou deux (cisaillement plan) parois rugueuses. 
La rugosité des parois est constituée de grains jointifs ayant les mêmes caractéristiques que les grains en écoulement. 
Des conditions aux limites périodiques sont appliquées selon la direction de l'écoulement : les écoulements sont simulés à 
l'intérieur d'une fenêtre de longueur $L$ (en général $40\times d$).

Une fois obtenu un état d'écoulement stationnaire, on mesure les profils moyens de compacité, taux de cisaillement, et contraintes 
(pression $P$ et contrainte de cisaillement $S$, on observe que les contraintes normales sont égales).
Cependant le matériau est loin d'être homogène, 
et l'on s'intéresse aussi aux fluctuations des grandeurs précédentes, ainsi qu'à certaines grandeurs microstructurelles : nombre de coordination, 
mobilisation du frottement (proportion des contacts glissants, pour lesquels le frottement est entièrement mobilisé)...

\subsection{Cisaillement homogène}

Nous commençons par discuter la géométrie de cisaillement la plus simple possible, à savoir le cisaillement plan sans gravité, 
où la distribution des contraintes est homogène (Fig. 13) [61, 75]. Le matériau est cisaillé entre deux parois parallèles distantes de 
$H$ (de $20$ à $100d$). Une des parois est fixée et l'autre se déplace à une vitesse imposée $V$. Par ailleurs la pression $P$ est contrôlée 
en autorisant la dilatation du matériau. À chaque pas de temps, la vitesse normale $v_n$ de la paroi mobile est déterminée à partir de la
 force normale $N$ exercée par les grains sur la paroi de la façon suivante : $v_n = (PL-N)/g_p$, où $g_p$ est un coefficient d'amortissement 
visqueux, de sorte que l'équilibre est obtenu quand $P = N/L$. Le contrôle de la pression introduit un nombre sans dimension  
$g_p/\sqrt{mK_N}$ en plus de 
ceux décrits au §2.3.4. Ce nombre est toujours petit dans nos simulations, ce qui signifie que l'échelle de temps des fluctuations de $H$ est 
imposée par le matériau et non par la paroi, et que la paroi <<~colle~>> au matériau.  
\begin{figure}[!htbp]
\bc
\includegraphics[height=6cm]{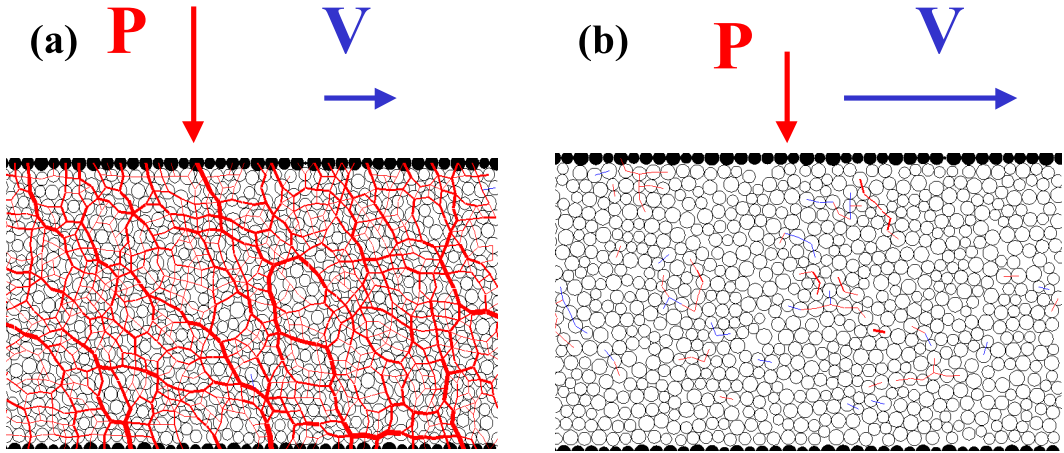}
\caption{Cisaillement plan. Réseau de contacts 
(a) Régime quasi-statique ($I = 10^{-2}$) ; 
(b) Régime dynamique ($I = 0.2$).
\label{fig13}
}
\ec
\end{figure}
Nous avons montré que l'état stationnaire est indépendant de la préparation du système (état lâche par dépôt aléatoire, 
ou dense par compaction cyclique sans frottement), et identique à celui de simulations à volume fixé [75]. Sauf dans les systèmes 
monodisperses et dans le cas des grains mous, on observe une absence de localisation. Cette géométrie modèle permet donc d'obtenir 
un matériau homogène, dont on contrôle bien l'état de cisaillement (pression $P$ et taux de cisaillement $\dot \gamma$).
Les écoulements stationnaires sont caractérisés par deux grandeurs macroscopiques moyennées dans l'espace
(dans la région centrale de la cellule de cisaillement, en excluant les cinq couches proches des parois)
et dans le temps : la compacité $\Phi$  et le frottement effectif $\mu^* = S/P$, rapport de la
contrainte de cisaillement sur la pression.

\subsubsection{ Grains non cohésifs}

L'analyse dimensionnelle a montré que l'état de cisaillement est caractérisé par le nombre d'inertie $I=\dot \gamma\sqrt{m/Pd}$ . 
Le régime quasi-statique ($I < 0.01$) est caractérisé par un réseau dense de contacts maintenus (Fig. 13a). Il correspond à l'état 
critique en mécanique des sols (cf. §3) et est caractérisé par une compacité maximale $\Phi _m$ et un frottement interne $\mu_S^*$. 
Lorsque $I$ augmente, c'est à dire lorsque le taux de cisaillement augmente et/ou la pression diminue, le milieu se dilate légèrement, 
le nombre de coordination diminue et la proportion de collisions augmente jusqu'à un régime dynamique purement collisionnel ($I>0.2$) (Fig. 13b). 
La transition entre les deux régimes (régime intermédiaire :  $0.01 < I < 0.2$) se fait de manière progressive. Corrélativement, 
l'évolution de la mobilisation du frottement conduit à une augmentation du frottement effectif. 
\begin{figure}[!htbp]
\bc
\includegraphics[height=6cm]{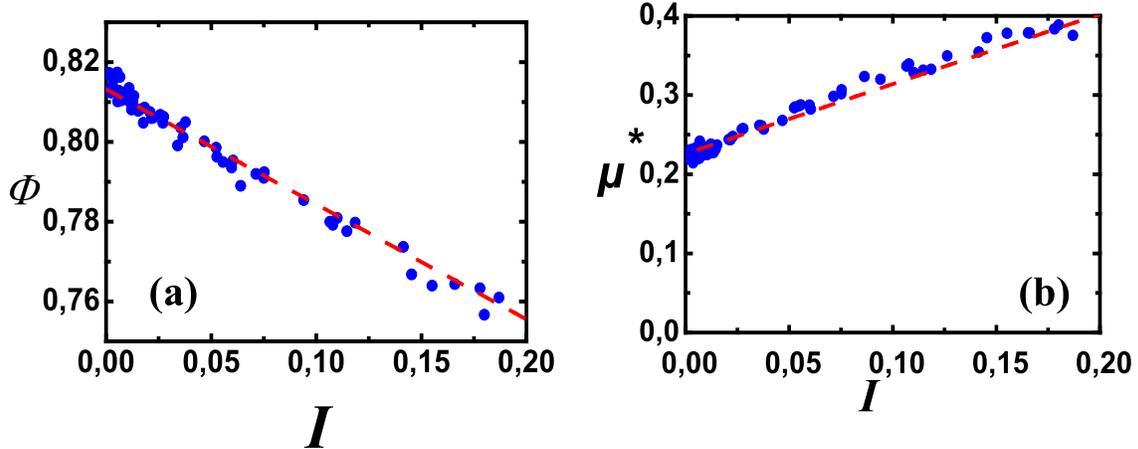}
\caption{Cisaillement plan ($\mu\ne 0$)- en rouge ajustement linéaire (11).
(a) Loi de dilatance ; 
(b) Loi de frottement.
\label{fig14}
}
\ec
\end{figure}
Les variations en fonction de $I$ de la 
compacité (<<~loi de dilatance~>>), du frottement effectif (<<~loi de frottement~>>), du nombre de coordination $z$ et de la mobilisation du 
frottement $M$, sont représentées sur les Fig. 14 et 15. On retiendra les dépendances simples suivantes (avec $a \simeq  1$ et $b \simeq 0,3$) :   
\be
\begin{aligned}
\Phi(I)&=\Phi_m-bI\\
\mu^*(I)&=\mu_S^*+aI\\
\end{aligned}
\label{eqn:11}
\ee
Ces deux lois permettent d'identifier la loi de comportement du matériau dans le régime intermédiaire, ce qui constitue une information 
précieuse dans les débats actuels (modèles hydrodynamiques inspirés de la transition vitreuse, frictionnel-collisionnel...[44, 61, 73, 74]). 
Elle est de type viscoplastique, avec un seuil d'écoulement de Coulomb et des contraintes visqueuses, dont l'origine est à chercher 
dans la réorganisation du réseau de contact, et qui dépendent du taux de cisaillement au carré et divergent près de la compacité maximale  : 
\be
\begin{aligned}
P&=b^2\frac{m\dot \gamma^2}{(\Phi_m-\Phi)^2}\\
S&=\mu_S^*P+ab\frac{m\dot \gamma^2}{(\Phi_m-\Phi)}\\
\end{aligned}
\label{eqn:12}
\ee
\begin{figure}[!htbp]
\bc
\includegraphics[width=16cm]{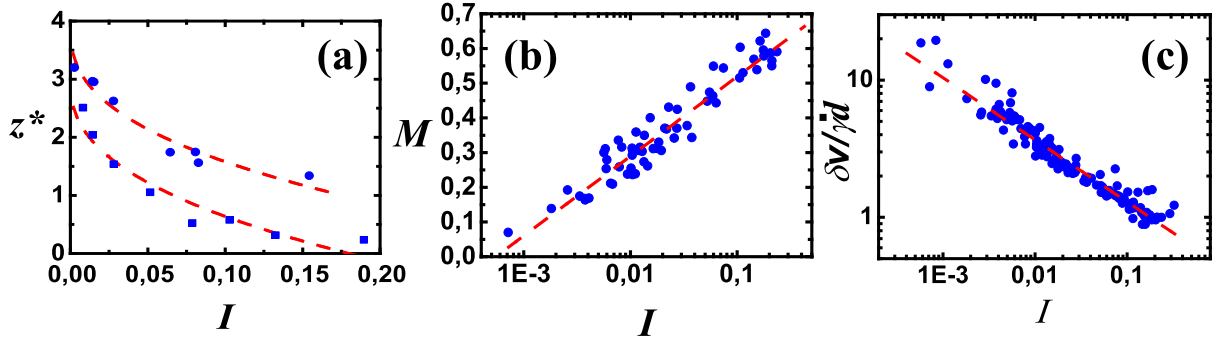}
\caption{Cisaillement plan - Aspects microstructurels
(a) Nombre de coordination pour $\mu =0$ (disques bleus, $e_N=0.1$ - carrés bleus : $e_N=0.9$) ;
(b) Mobilisation du frottement ;
(c) Fluctuations relatives de la vitesse de translation.
\label{fig15}
}
\ec
\end{figure}
Dans le régime dynamique, susceptible d'être décrit par la théorie cinétique des gaz denses [10], la loi de comportement dépend du 
coefficient de restitution (Fig. 16a). En revanche, dans le régime intermédiaire, les lois de dilatance et de frottement sont à peu près 
indépendantes des caractéristiques mécaniques des grains (coefficients de restitution et de frottement, rigidité). On doit seulement distinguer 
le cas des grains sans frottement, où la rotation des grains ne joue alors plus de rôle (Fig. 16a): la loi de frottement garde la même allure 
mais est décalée vers des valeurs de frottement plus faibles ($\mu_S^*$ diminue de $0.2$ à $0.1$).
Par ailleurs, la compacité maximale $\Phi_m$ n'a pas une 
signification purement géométrique, mais diminue lorsque le frottement microscopique $\mu$  augmente  (Fig. 16b - insert). La représentation du 
frottement effectif en fonction de la compacité (Fig. 16b) met en évidence une courbe maîtresse regroupant l'ensemble des données. Il est 
remarquable qu'une petite variation de compacité (de l'ordre de $10\%$) suffise à induire une variation du frottement d'un facteur 4. 
\begin{figure}[!htbp]
\bc
\includegraphics[width=16cm]{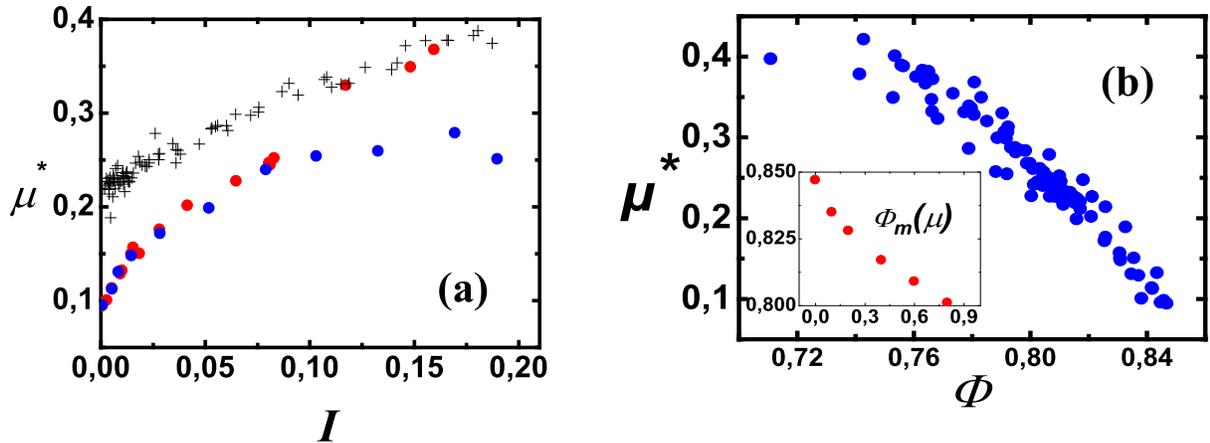}
\caption{Cisaillement plan : influence des caractéristiques mécaniques des grains. 
(a) Influence de $\mu$  et de $e_N$. Les croix correspondent au cas $\mu \ne 0$, les disques rouges au cas  $\mu =0$ et  $e_N =0.1$,
les disques bleus au cas  $\mu =0$ et  $e_N =0.9$.
(b) Loi de frottement en fonction de la compacité. En insert, dépendance de la compacité maximale en fonction de $\mu$ .
\label{fig16}
}
\ec
\end{figure}

Dans le régime quasi-statique, lorsque $I\to 0$, les fluctuations relatives de vitesse de translation $\delta v/\dot \gamma$
augmentent, selon une loi de puissance (Fig. 15c), tandis que les fluctuations relatives des contraintes diminuent.
L'écoulement semble intermittent pour $I < 0.001$, alors qu'il est 
continu et stationnaire pour les $I$ plus grands. Ces intermittences semblent associées à des mouvements corrélés de blocs, observés 
expérimentalement et numériquement [76-78]. Ils ont motivé l'élaboration de plusieurs modèles non-locaux [44, 79-81]. 

\subsubsection{Grains cohésifs}

L'étude de l'influence de la cohésion a été menée à travers le modèle simple présenté au §2.1.4 [82, 83]. L'analyse dimensionnelle 
(§2.3.4) a fait ressortir deux nombres sans dimension qui contrôlent l'état de cisaillement ($I$) et l'intensité de la cohésion ($\eta$). 
Pour une cohésion assez faible ($\eta < 12$), on observe des écoulements stationnaires. La loi de dilatance reste qualitativement la même 
que celle observée avec des grains secs (11), mais la compacité maximale diminue fortement avec la cohésion (Fig. 17a). 
La loi de frottement (11) n'est pas modifiée. Cependant, pour une cohésion assez forte ($\eta < 12$), on n'observe plus 
d'écoulements stationnaires dans la gamme de $I$ explorée. Le frottement augmente alors notablement (Fig. 17b - courbe rouge), 
ainsi que les fluctuations. 
 
On met donc en évidence la transition entre deux régimes d'écoulements, stationnaire (inertiel et/ou faible cohésion) et 
intermittent (quasi-statique et/ou forte inertie), dont l'origine se trouve à l'échelle des contacts entre grains. 
La valeur critique de cohésion obtenue est en effet de l'ordre de celle anticipée par l'analyse dimensionnelle du §2.3.4 sur 
des interactions binaires ($\eta \simeq4$). On observe que la cohésion provoque une diminution de la compacité, mais une augmentation du nombre 
de coordination, et une homogénéisation des directions de contact. C'est le signe d'une organisation des grains en amas 
compacts séparés par du vide. La proportion de vide supplémentaire ne dépend pas de l'état de cisaillement $I$ mais varie 
linéairement avec la cohésion. De plus, la cohésion augmente la durée de vie des contacts. Lorsque tous les contacts persistent, 
le milieu se déplace en blocs rigides qui se collent alternativement à l'une ou l'autre des parois : seules les zones proches des 
parois sont alors cisaillées.
\begin{figure}[!htbp]
\bc
\includegraphics[width=16cm]{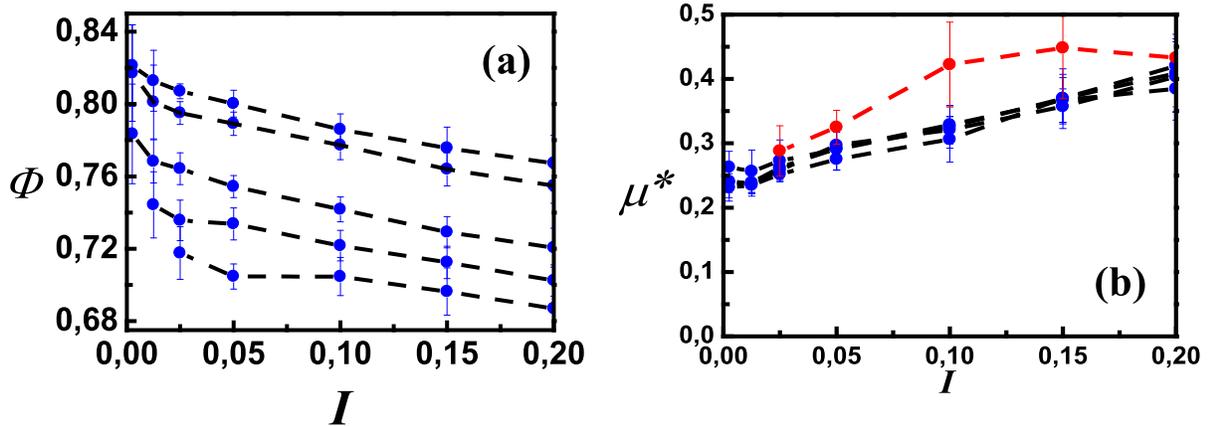}
\caption{Cisaillement plan - grains cohésifs ($\eta = 0$, $0.25$, $4$, $12$, $25$).
Les barres représentent les fluctuations dans l'épaisseur de l'écoulement.
(a) Loi de dilatance.
(b) Loi de frottement - en rouge $\eta = 25$.
\label{fig17}
}
\ec
\end{figure}

\subsection{Plan incliné}

L'étude du comportement à vitesse imposée en présence d'une hétérogénéité des contraintes (cisaillement plan avec gravité et 
cisaillement annulaire [61]) a montré que les lois de frottement et de dilatance restent qualitativement les mêmes. L'hétérogénéité des 
contraintes conduit à alors à une localisation du cisaillement. 

Considérons maintenant le cas des écoulements sur un plan incliné rugueux. Nous allons voir comment les simulations numériques discrètes, 
en complément d'expériences sur matériau modèle, ont permis de mesurer le comportement, en particulier le frottement effectif sur plan 
incliné [16, 61]. La connaissance de cette grandeur permet de prédire l'étalement d'une masse granulaire sur une pente, dans le cadre 
d'une description moyennée sur l'épaisseur [84, 85]. Le bon accord entre expériences et simulations numériques [85] est prometteur pour 
la prédiction de la propagation des écoulements géophysiques [86].

\subsubsection{Epaisseur d'arrêt}

Le blocage de l'écoulement d'une couche de grains sur un plan rugueux incliné dépend de l'inclinaison $\theta$
mais aussi de l'épaisseur $H$. 
On mesure ainsi une épaisseur d'arrêt Hstop(q) (Fig. 18a),
mise en évidence expérimentalement pour divers couples matériau-rugosité [74, 87], 
et décrite par : 
\be
H_{stop}(\theta)=B\frac{\theta_M-\theta}{\theta-\theta_m}
\label{eqn:hstop}
\ee
Les paramètres $\theta_m$ et $\theta_M$ (bornes pour un écoulement épais ou mince) et $B$ (longueur d'influence de la rugosité) 
dépendent du couple matériau-rugosité. Dans nos simulations en deux dimensions (Fig. 18a), $\theta_m=13.5$°,
$\theta_M = 35$° et $B/d\simeq 1.2$.
\begin{figure}[!htbp]
\bc
\includegraphics[width=16cm]{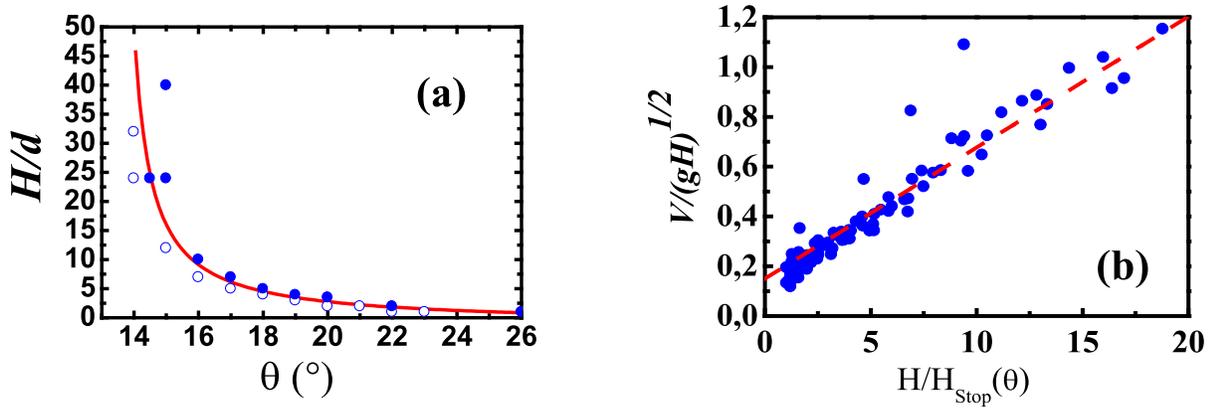}
\caption{Plan incliné.
(a) Épaisseur d'arrêt (courbe rouge). Les cercles bleus correspondent à l'arrêt de l'écoulement, les disques bleus à un écoulement stationnaire
(b) Loi d'écoulement. Les trois disques bleus qui s'écartent de la loi simple correspondent à des écoulements épais ($H = 40$).
\label{fig18}
}
\ec
\end{figure}
\subsubsection{Ecoulement stationnaire et frottement effectif}

Dans le régime d'écoulement stationnaire au-dessus du seuil, on a mesuré les profils de compacité et de vitesse (Fig. 19a). 
La mesure expérimentale des dépendances de la vitesse moyenne $V$ en fonction de $H$ et de $\theta$ 
(<<~loi d'écoulement~>>) a montré que $V$  s'exprime très simplement à l'aide de l'épaisseur d'arrêt 
$Hstop(\theta)$, sous la forme d'une relation, entre deux nombres sans dimension, 
relatifs l'un à la vitesse et l'autre à l'épaisseur [87] : 
\be
\frac{V(H,\theta)}{\sqrt{gH}}=\beta\frac{H}{H_{stop}(\theta)},
\label{eqn:poul}
\ee
avec $g$ la gravité et $b$  une constante dépendant du couple matériau-rugosité ($b \simeq 0.1$). La Fig. 18b 
montre une mesure par simulation numérique discrète.
\begin{figure}[!htbp]
\bc
\includegraphics[width=16cm]{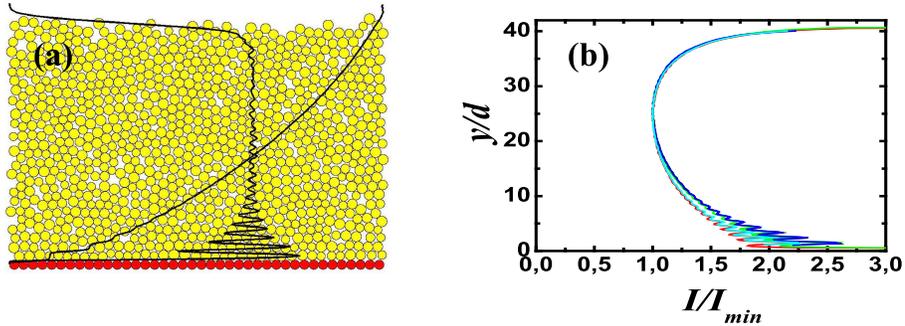}
\caption{Plan incliné 
(a) Image des grains en écoulement - Profils de compacité et de vitesse ($H=25$, $\theta =18$°)
(b) Profils normalisés du nombre inertiel pour différentes inclinaisons ($H=40$, $\theta$  entre $15$° et $21$°)
\label{fig19}
}
\ec
\end{figure}
Il est alors possible de remonter à la loi de frottement, c'est à dire à la dépendance du frottement effectif en fonction du nombre d'inertie. 
Pour un écoulement stationnaire uniforme, le frottement effectif est égal à l'inclinaison $\theta$ . Les profils du nombre inertiel (Fig. 19b) 
montre une augmentation de $I$  près de la surface libre (due à la fluidisation) et à proximité du socle (diminution de la viscosité effective 
due à la structuration du milieu en couches qui glissent plus facilement les unes sur les autres). La forme de ces profils est indépendante de 
l'inclinaison, mais l'amplitude diminue lorsque l'on se rapproche de l'arrêt. La variation de $I$ au centre de
l'écoulement en fonction de$\theta$ fournit la loi de frottement sur plan incliné, représentée sur la Fig. 20a.
On constate que celle-ci s'écarte de celle mise en évidence en 
cisaillement homogène. La loi d'échelle (14) conduit à (avec $a = 2B/5\beta$) : 
\be
\mu^*(I)=\frac{\theta_m+\alpha\theta_MI}{1+\alpha I}.
\label{eqn:15}
\ee
\subsubsection{Blocage}

Selon la loi de frottement (15), l'écoulement devrait s'arrêter progressivement pour $\theta = \theta_m$. 
En fait, on observe un arrêt brutal lorsque $\theta < \theta_{stop}(H)$, ce qui correspond à une valeur critique de 
$I$ dépendant de $H$ $(I_{stop}(H)=5\beta d/2H$) [61]. Ceci se traduit par une rupture dans la loi de frottement, 
représentée par les traits bleus sur la Fig. 20b. 
\begin{figure}[!htbp]
\bc
\includegraphics[width=16cm]{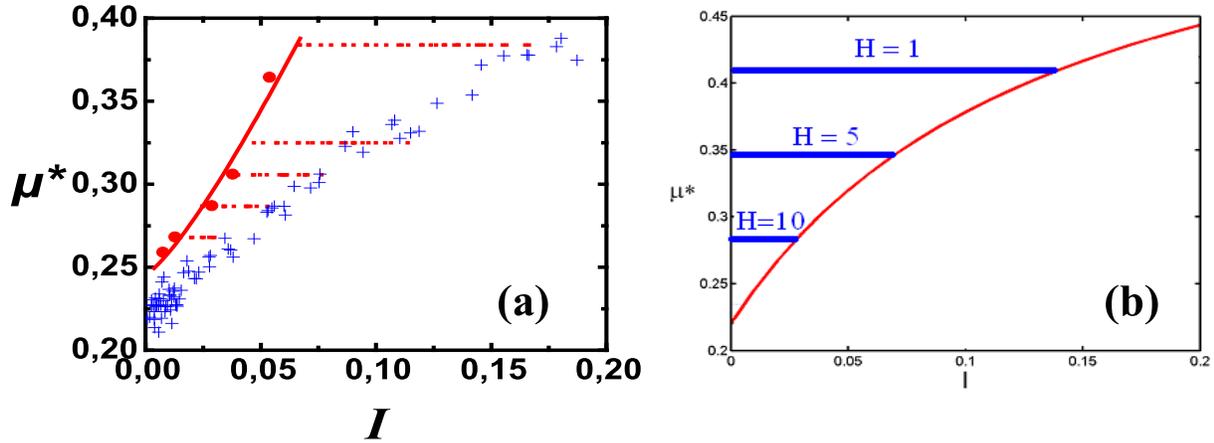}
\caption{Loi de frottement sur plan incliné.
(a) Les croix bleues correspondent au cisaillement plan, les disques rouges (dynamique des contacts) et la courbe rouge (dynamique moléculaire)
sont déduites de la mesure du nombre d'inertie au coeur de la couche.
Les courbes en pointillé rouge correspondent aux profils du nombre d'inertie. 
(b) Prédiction d'un modèle de piégeage - Effet de l'épaisseur sur l'arrêt de l'écoulement.
\label{fig20}
}
\ec
\end{figure}
Ceci est à rapprocher de l'apparition d'intermittences et de 
mouvements de blocs en cisaillement plan à vitesse imposée lorsque $I \to 0$. Lorsque l'écoulement ralentit, 
la taille des blocs augmente et lorsque celle-ci devient comparable à l'épaisseur de la couche en écoulement, 
le matériau s'arrête. Un modèle phénoménologique a été proposé [61,  88], fondé sur un effet de piégeage des grains initié par 
la paroi rugueuse, en bon accord avec les observations du comportement d'une couche de grains sur un plan incliné au voisinage 
de l'arrêt (épaisseur d'arrêt, loi d'écoulement et de frottement, arrêt brutal) (Fig. 20b). La compréhension de ce type d'écoulement 
nécessite donc de prendre en compte l'interaction du matériau avec la paroi rugueuse. 
	
Par ailleurs, en diminuant soudainement l'inclinaison à partir d'un écoulement stationnaire, 
les simulations numériques ont permis de mesurer l'évolution du réseau de contact lors de la transition fluide-solide. 
Trois grandeurs micromécaniques (nombre de coordination et mobilisation du frottement, fluctuations relatives de la vitesse de translation) 
manifestent une transition brutale entre une valeur dans l'état fluide et une valeur dans l'état solide (Fig. 21). 
Ainsi, la mobilisation du frottement chute brusquement à 0, tandis que le nombre de coordination et les fluctuations relatives de 
vitesse augmentent notablement. Le blocage est donc associé à une transition microstructurelle massive.
\begin{figure}[!htbp]
\bc
\includegraphics[width=16cm]{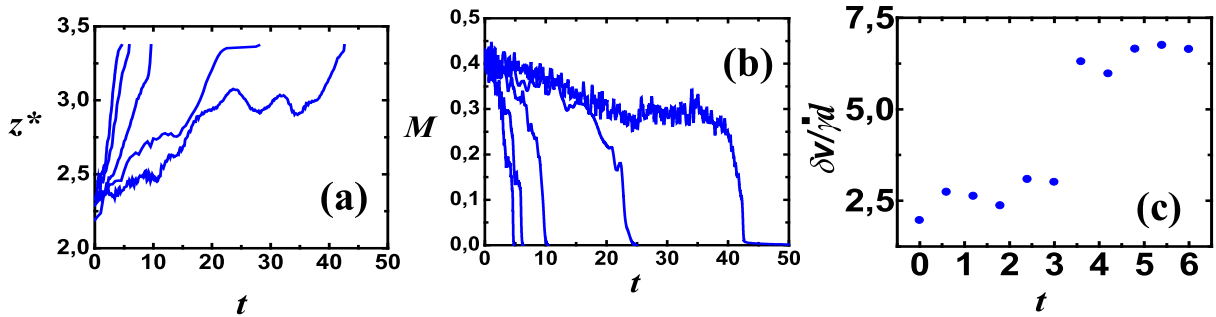}
\caption{Transition de blocage - Evolution temporelle à partir d'un écoulement stationnaire ($H=20$, $\theta=17$°)
lors d'une variation brutale de l'inclinaison
(a) Nombre de coordination ($\Delta \theta$ entre $3$° et $17$°)
(b) Mobilisation du frottement ($\Delta \theta$ entre $3$° et $17$°)
(c) Fluctuations relatives de la vitesse de translation ($\Delta \theta = 12$°)
\label{fig21}
}
\ec
\end{figure}

\subsection{Conclusion}

	La simulation numérique discrète a donc permis de progresser dans la connaissance de la loi de comportement 
des écoulements granulaires denses et des mécanismes de blocage. Les informations décrites doivent être confirmées dans 
d'autres géométries : cisaillement plan avec gravité, cisaillement annulaire, conduite verticale, socle meuble [61, 73, 74]. 
Elles devraient permettre de comprendre le rôle des corrélations et d'étudier 
les comportements particuliers au voisinage des parois et leur interaction avec le comportement en volume.

\section{Conclusion}

La simulation numérique discrète contribue à améliorer la compréhension des comportements macroscopiques des matériaux granulaires, 
en permettant de mesurer avec précision l'influence des divers paramètres. C'est ce que tente de résumer le Tableau~\ref{tab:param}
(dont le contenu appellerait certaines nuances et investigations complémentaires). En plus des cinq nombres sans dimensions 
identifiés au §2, nous avons introduit l'état initial dont la géométrie est déterminante tant qu'un régime de déformation 
stationnaire n'est pas atteint. On s'attend aussi bien sûr à une influence sensible de la granulométrie et de la forme des grains,
 qui reste à étudier. Les études évoquées au §3 et 4 montrent que le passage du modèle de contact à 
l'échelle du grain aux comportements collectifs présente des aspects inattendus. 
\begin{table}
\caption{Influence de la préparation et des paramètres sans dimension identifiés au §2 sur les comportements mécaniques. (O=Oui, N=Non)}
\centering
\small
\begin{tabular}{|c||c|c|c|c|c|c|}  \cline{1-7}
 &{\scriptsize État initial} & $I$ & $\kappa$ ({\scriptsize $>10^{-3}$})&  $e$ ou $\zeta$& $\mu$& $\eta$\\
\hline
\hline
{\scriptsize Préparation} &O&&N&O&O&O\\
\hline
{\scriptsize Très petites déformations, $\epsilon \sim 10^{-5}$}&O&N&O&N&O&O ?\\
\hline
{\scriptsize Petites déformations, $\epsilon \sim 10^{-2}$}&O&N {(\scriptsize $I<10^{-3}$)}&N&N {\scriptsize ($I<10^{-3}$)}&O&O ?\\
\hline
{\scriptsize État critique, $\epsilon \sim 1$} &N&N {\scriptsize ($I<10^{-4}$ ?)}&N&N {\scriptsize($I<10^{-4}$ ?)}&O& O ?\\
\hline
{\scriptsize Écoulements denses, $10^{-3}<I<0.1$} &N&O&N&N&O&O\\
\hline
{\scriptsize Régime collisionnel, $I>0.1$} &N&O&N ?&O&O& ?\\
\hline
\end{tabular}
\label{tab:param}
\normalsize
\end{table}
C'est ainsi que la rhéologie est indépendante 
de la déformation des contacts quand la déformation macroscopique dépasse  $10^{-3}$, ou que le frottement interne est indépendant 
du frottement intergranulaire dans un écoulement stationnaire dense. En revanche, on a mis en évidence l'influence de la 
géométrie de l'assemblage sur le comportement en régime quasi-statique, ainsi que le rôle dominant dans les écoulements denses 
du paramètre d'inertie, qui décrit l'état de cisaillement. Les travaux actuels mettent ainsi en lumière l'importance des mécanismes collectifs 
à l'origine des comportements macroscopiques : après l'intérêt récent pour les distributions hétérogènes des efforts, 
l'attention se porte maintenant vers l'analyse des mouvements et des déformations corrélés. Dans les écoulements denses, 
il semble que le paramètre d'inertie détermine l'échelle caractéristique de ces corrélations. Dans les assemblages solides, 
de faibles incréments de déformation ont pour origine une succession de ruptures et de réparations du réseau des contacts. 
Cette identification des mécanismes physiques responsables de la déformation et de l'écoulement est indispensable à  
l'élaboration des modèles rhéologiques pour les matériaux granulaires.

\noindent
{\large {\bf Remerciements~:}}
Nous souhaitons remercier Gaël Combe, Frédéric da Cruz, Sacha Emam, Michaël Prochnow et Pierre Rognon 
qui ont contribué à certains des résultats présentés ci-dessus, Michel Jean et Jean-Jacques Moreau, 
qui nous ont initiés à la méthode de dynamique des contacts, et de nombreux collègues du LMSGC, du CERMES, 
du LPMDI, de l'ENTPE, du Groupe de Recherche du CNRS sur les Milieux Divisés et de l'ancien réseau de laboratoires GEO 
pour des contacts et des discussions sur le comportement mécanique des matériaux granulaires et sa modélisation numérique.

\section*{Bibliographie}
\bn
\im
DURAN J., {\em Sables, poudres et grains}, Eyrolles Sciences, Paris, {\bf 1999}.
\im
NEDDERMAN R.M., {\em Statics and kinematics of granular materials}, Cambridge University Press, {\bf 1992}.
\im
HERRMANN H.J, HOVI J.-P et LUDING  S. (eds.), {\em Physics of dry granular media}, Kluwer Academic, {\bf 1998}.
\im
ODA M. et IWASHITA K. (eds), {\em Mechanics of granular materials, an introduction}. Balkema, Rotterdam, {\bf 1999}. 
\im
CHEVOIR F. et ROUX J-N. (eds), {\em Colloque physique et mécanique des matériaux granulaires}, 
Collection Actes des journées scientifiques du LCPC, {\bf 2000}.
\im
KISHINO Y. (ed.), {\em Powders and grains},  Swets \& Zeitlinger, Lisse, {\bf 2001}.
\im
DURAN J et BOUCHAUD J-Ph. (eds), {\em Physics of granular media}, Comptes Rendus de l'Académie des Sciences Physique, tome 3, n°2, mars {\bf 2002}.
\im
WOOD D.M., {\em Soil behaviour and critical state soil mechanics}, Cambridge University Press,  {\bf 1990}.
\im
COUSSOT Ph. et ANCEY C.,{\em  Rhéologie des pâtes et des suspensions}, EDP Sciences, Les Ulis, {\bf 1999}.
\im
AZANZA E., CHEVOIR F. et MOUCHERONT P., Experimental study of collisional granular flows down an inclined plane, 
 {\em Journal of Fluid Mechanics}, 400, {\bf 1999}, pp. 199-227.
\im
CAMBOU B. et JEAN M. (dir), {\em Micromécanique des matériaux granulaires}. Hermès, Paris, {\bf 2001}.
\im
BOSSIS G. et BRADY J.F., Stokesian dynamics, {\em  Annual Review of Fluid Mechanics}, 20, {\bf  1988}, pp. 111-157.
\im
MAILLARD S., DE LA ROCHE C., HAMMOUM F., GAILLET L., DAOUBEN E. et SUCH C., 
Comportement à la rupture du bitume en film mince sous chargement répété - 
Approche par des méthodes de contrôle non destructif, 
 {\em Actes des Journées Sciences de l'Ingénieur du Réseau des Laboratoires des Ponts et Chaussées}, {\bf 2003}, pp. 239-245. 
\im
DE LARRARD F.,  {\em Structure granulaire et formulation des bétons}, 
Etudes et Recherches des Laboratoires des Ponts et Chaussées (OA34), {\bf 2000}.
\im
SILBERT L., ERTAS D., GREST G.S., HALSEY T.C.. et LEVINE D., Geometry of frictionless and 
frictional sphere packings, Physical Review E, 65, {\bf 2002}, pp. 031304
\im
SILBERT L., ERTAS D., GREST G.S., HALSEY T.C., LEVINE D. et PLIMPTON S.J.,
 Granular flow down an inclined plane.  {\em Physical Review E}, 64, {\bf 2001}, pp. 051302.
\im
PÖSCHEL T. et BUCHHOLTZ V., Molecular dynamics of arbitrarily shaped particles,  {\em Journal de Physique I}, 5, {\bf 1995}, pp. 1431-1455.
\im
PETIT D., PRADEL F., FERRER G. et MEIMON Y., Shape effect of grain in a granular flow, in [6], pp. 425-428.
\im
OVIEDO X., SAB K. et GAUTIER P-E., Etude du comportement du ballast ferroviaire par un modèle micromécanique, in [5], pp. 339-344.
\im
DIMNET E.,  {\em Mouvement et collisions de solides rigides ou déformables}, Thèse de l'Ecole Nationale des Ponts et Chaussées, {\bf 2002}.
\im
VU-QUOC L., ZHANG X. et WALTON O.R., A 3D discrete-element method for dry granular flows of ellipsoidal particles, 
 {\em Computer Methods in Applied Mechanics Engineering}, 187, {\bf 2000}, pp. 483-528.
\im
SCHÄFER, J., DIPPEL S. et WOLF D., Force schemes in simulations of granular materials,  {\em Journal de Physique I}, 6, {\bf 1996}, pp. 5-20.
\im
JOHNSON K.L.,  {\em Contact Mechanics}, Cambridge University Press, {\bf  1985}.
\im
ELATA D. et BERRYMAN J.G., 
Contact force-displacement laws and the mechanical behavior of random packs of identical spheres, 
 {\em Mechanics of Materials}, 24, {\bf 1996}, pp. 229-240.
\im
BRILLIANTOV N.V., SPAHN F., HERTZSCH J.M. et PÖSCHEL T., Model for collisions in granular gases,  
{\em Physical Review E}, 53, {\bf 1996}, pp. 5382-5392.
\im
WALTON O.L. et BRAUN R.L., Viscosity, granular temperature,
 and stress calculations for shearing assemblies of inelastic, frictional disks,  {\em Journal of Rheology}, 30, {\bf  1986}, pp. 949-980.
\im
PITOIS O., {\em  Assemblée de grains lubrifiés : Elaboration d'un système modèle expérimental et étude de la loi de contact}, 
Thèse de doctorat de l'Ecole Nationale des Ponts et Chaussées, {\bf 1999}
\im
CHATEAU X., MOUCHERONT P. et PITOIS O., Micromechanics of unsaturated granular media, 
 {\em ASCE Journal of Engineering Mechanics}, 128, {\bf 2002}, pp. 856-863.
\im
MAUGIS D.,  {\em Contact, adhesion and rupture of elastic solids}, Springer Verlag, Berlin, {\bf 2000}.
\im
GREENWOOD J.A. et JOHNSON K.L., An alternative to the Maugis model of adhesion between elastic spheres, 
 {\em Journal of Physics D}, 31, {\bf 1998}, pp. 2379-3290.
\im
THORNTON C. et YIN K.K., Impact of elastic spheres with and without adhesion, {\em  Powder Technology}, 65, {\bf  1991}, pp. 153-166.
\im
PREECHAWUTTIPONG, I.,  {\em Modélisation du comportement mécanique de milieux granulaires cohésifs}, 
Thèse de doctorat de l'Université Montpellier II, {\bf 2002}.
\im
MATUTTIS H.G. et SCHINNER A., Particule simulation of cohesive granular materials, {\em  International Journal of Modern Physics C}, 
12, {\bf 2001}, pp.1011-1021.
\im
DELENNE J-Y., SAID EL YOUSSOUFI M. et BENET J-C., Comportement mécanique et rupture de milieux granulaires cohésifs, 
 {\em Comptes-Rendus de l'Académie des Sciences, Mécanique}, 330, {\bf 2002}, pp. 1-8. 
\im
IWASHITA K. et ODA M., Rotational resistance at contacts in the simulation of shear band development by DEM, 
 {\em ASCE Journal of Engineering Mechanics}, 124, {\bf 1998}, pp. 285-292.
\im
MAW N., BARBER J-R. et FAWCETT J.N., The oblique impact of elastic spheres, {\em  Wear} 38, {\bf  1976}, pp. 101-114.
\im
WOLF D.E., Modeling and computer simulation of granular media, in HOFFMANN K.H. et SCHREIBER M. (eds), {\em  Computational physics. 
Selected methods, simple exercises, serious applications}, Springer-Verlag, Berlin, {\bf 1996}.
\im
HERRMANN H.J. et LUDING S., Modeling granular media on the computer,  {\em Continuum Mechanics  and Thermodynamics}, 10, {\bf 1998}, pp. 189-231.
\im
ALLEN M.P. et TILDESLEY  D.J.,  {\em Computer Simulation of Liquids}, Oxford University Press, {\bf  1987}.
\im
RAPAPORT D.C., {\em  The art of molecular dynamics simulation}, Cambridge University Press, {\bf 1995}.
\im
FRENKEL D. et SMIT B., {\em  Understanding molecular simulation}, Academic Press, {\bf 1996}.
\im
ROUX J-N., La simulation numérique discrète dans l'étude des matériaux : 
l'exemple des assemblages de sphères.  
{\em Actes des Journées Sciences de l'Ingénieur du Réseau des Laboratoires des Ponts et Chaussées}, {\bf 2003}, pp. 525-534.
\im
CUNDALL P.A. et STRACK O.D.L., A discrete numerical model for granular assemblies,  {\em Géotechnique} 29, {\bf  1979}, pp. 47-65.
\im
PROCHNOW M., {\em  Ecoulements denses de grains secs},  Thèse de l'Ecole Nationale des Ponts et Chaussées, {\bf 2002}.
\im
MOREAU J.-J. et JEAN M., Unilaterality and dry friction in the dynamics of rigid body collections, in CURNIER A., 
 {\em Proceedings of Contact Mechanics International Symposium}, {\bf 1992}, pp 31-48. 
\im
MOREAU J-J., Some numerical methods in multibody dynamics : application to granular materials, 
 {\em European  Journal of Mechanics }A, 13, {\bf 1994}, pp. 93-114.
\im
JEAN M., The non-smooth contact dynamics method,  {\em Computer methods in applied mechanics and engineering}, 177, {\bf 1999}, pp. 235-257.
\im
RADJAI F. , BRENDEL L. et ROUX S., Nonsmoothness, indeterminacy and friction in two-dimensional arrays of rigid particles, 
 {\em Physical Review E}, 54, {\bf 1996}, 861-873.
\im
MOREAU J.J., Numerical investigation of shear zones in granular material, in WOLF D.E. et GRASSBERGER P. (eds), 
 {\em Friction, Arching, Contact Dynamics}, World Scientific, Singapour, {\bf 1997}, pp. 233-247.
\im
STARON L., VILOTTE J-P. et RADJAI F., Pre-avalanche instabilities in a tilted granular pile,  {\em Physical Review Letters}, 89, {\bf 2002}, pp. 204302.
\im
RADJAI F. et ROUX S., Contact dynamics study of 3D granular media : critical states and relevant internal variables, in 
 {\em The physics of granular media}, HINRICHSEN H. et WOLF D.E (eds), Wiley-Vch, Berlin, {\bf 2003}. 
\im
MOREAU J.J., Numerical experiments in granular dynamics : vibration-induced size segregation, in  {\em Contact Mechanics}, 
RAOUS M. (ed), Plenum Press, New York, {\bf 1995}, pp. 347-358.
\im
AZANZA E.,  {\em Ecoulements granulaires bidimensionnels sur plan incliné},  
Etudes et Recherches des Laboratoires des Ponts et Chaussées (SI5), {\bf 1998}. 
\im
KADAU D., BARTELS G., BRENDEL L. et WOLF D.E., 
Contact dynamics simulations of compacting cohesive granular systems, {\em  Computer Physics Communications}, 147, {\bf 2002}, pp. 190-193.  
\im
BOURADA-BENYAMINA N., {\em  Etude du comportement des milieux granulaires par homogénéisation périodique}, 
Thèse de l'Ecole Nationale des Ponts et Chaussées, {\bf 1999}.
\im
COMBE G.,  {\em Mécanique des matériaux granulaires et origines microscopiques de la déformation},  
Etudes et Recherches des Laboratoires des Ponts et Chaussées (SI8), {\bf 2002}.
\im
ROUX J.-N. et COMBE G., Quasistatic rheology and the origins of strain, in [7] pp. 131-140.
\im
JEFFERSON G, HARITOS G.K et MCMEEKING R.M., The elastic response of a cohesive aggregate - 
a discrete element model with couples particle interaction,   {\em Journal of the Mechanics and Physics of Solids}, 50, {\bf 2002},  pp. 2539-2575.
\im
BIDEAU D. et HANSEN A. (eds), {\em  Disorder and Granular Media}, North-Holland, Amsterdam, {\bf 1993}. 
\im
THORNTON C., Numerical simulations of deviatoric shear deformation of granular media, {\em Géotechnique}, 50, {\bf 2000}, pp. 43-53.
\im
DA CRUZ F., {\em Frottement et blocage des écoulements de grains secs.} Thèse de l'Ecole Nationale des Ponts et Chaussées, {\bf 2004}.
\im
DANTU P., Contribution à l'étude mécanique et géométrique des milieux pulvérulents, in 
 {\em Proceedings of the 4th International Conference on Soil Mechanics and Foundation Engineering}, 
tome 1, Butterworth, London, {\bf  1957}, pp. 144-148. 
\im
SAB K. (ed),  {\em Approches multiéchelles pour les matériaux et les structures du génie civil}, 
Collection Actes des journées scientifiques du LCPC, {\bf 2000}.
\im
BENAHMED N., {\em Comportement mécanique d'un sable sous cisaillement monotone et cyclique. 
Applications à la liquéfaction et à la mobilité cyclique.} Thèse de l'Ecole Nationale des Ponts et Chaussées, {\bf 2001}.
\im
EMAM S., thèse de l'Ecole Nationale des Ponts et Chaussées en préparation, {\bf  2005}.
\im
CUMBERLAND D.J. et CRAWFORD R.J., {\em The Packing of Particles}, Elsevier, {\bf  1987}.
\im
DOMENICO S.N., Elastic properties of unconsolidated porous sand reservoirs, {\em Geophysics}, 42, {\bf  1977}, pp. 1339-1368. 
\im
JIA X. et MILLS P., Sound propagation in dense granular materials, in [6] pp. 105-112. 
Ces auteurs sont également remerciés pour avoir aimablement communiqué des données non publiées.
\im
GEOFFROY H., DI BENEDETTO H., DUTTINE A. et SAUZÉAT C., Dynamic and cyclic loadings on sands, results and modeling 
for general stress-strain conditions pp. 353-363, in DI BENEDETTO H., DOANH T., GEOFFROY H. et SAUZÉAT C. (eds), 
 {\em Deformation characteristics of geomaterials}, Swets and Zeitlinger, Lisse, {\bf 2003}. 
\im
COMBE G. et ROUX J-N., Discrete numerical simulation, quasistatic deformation and the origins of 
strain in granular materials, pp.1071-1078, in DI BENEDETTO H., DOANH T., GEOFFROY H. et SAUZÉAT C. (eds), 
 {\em Deformation characteristics of geomaterials}, Swets and Zeitlinger, Lisse, {\bf 2003}.
\im
LIU A. et NAGEL S. (eds), {\em Jamming and rheology}, Taylor and Francis, {\bf 2001}.  
\im
DA CRUZ F., CHEVOIR F., BONN D. et COUSSOT Ph., Viscosity bifurcation in granular materials, foams and emulsions, 
 {\em Physical Review }E, 66, {\bf 2002}, 051305.
\im
POULIQUEN O. et CHEVOIR F., Dense flows of dry granular materials in [7], pp. 163-175.
\im
GDR MIDI, On dense granular flows, accepté pour publication dans {\em European Journal of Physics}, {\bf 2004}. 
\im
DA CRUZ F., CHEVOIR F., ROUX J-N. et IORDANOFF I., Macroscopic friction of dry granular materials, in DALMAZ G. (ed) 
 {\em Actes du 30ème Symposium Leeds-Lyon de Tribologie}, {\bf 2003}.
\im
CHAMBON G., SCHMITTBUHL J., CORFDIR A, VILOTTE J.P. et ROUX, S., Shear with comminution of a granular material : 
Microscopic deformations outside the shear band, {\em Physical Review }E, 68, {\bf 2003}, pp. 011304.
\im
BONAMY D., DAVIAUD F., LAURENT L., BONETTI M. et BOUCHAUD J.P., 
Multiscale clustering in granular surface flows,  {\em Physical Review Letters}, 89, {\bf 2002}, 034301.
\im
RADJAI F. et ROUX S., Turbulent-like fluctuations in quasi-static flow of granular media, 
{\em Physical Review Letters}, 89, {\bf 2002}, pp. 064302.
\im
POULIQUEN O., FORTERRE Y. et LEDIZES S., Dense granular flows down incline as a self activated process, 
 {\em Advances in Complex systems}, 4, {\bf 2001}, pp. 441-450.
\im
ERTAS D. et HALSEY T.C., Granular gravitational collapse and chute flow, {\em Europhysics Letters}, 60, {\bf 2002}, pp. 931.
\im
MILLS P., LOGGIA D. et TIXIER M., Model for a stationary dense granular flow along an inclined wall,
{\em Europhysics Letters}, 45, {\bf 1999}, pp. 733-738.
\im
ROGNON P., DA CRUZ F., EMAM S., ROUX J-N. et CHEVOIR F., Rhéologie des matériaux granulaires cohésifs : 
simulation numérique du cisaillement plan, {\em Colloque Science et Technologie des Poudres}, {\bf 2004}.
\im
IORDANOFF I., SÈVE B. et BERTHIER Y., Solid third body analysis using a discrete approach : 
influence of adhesion and particle size on the macroscopic behavior of the contact, {\em ASME Journal of Tribology}, 124, {\bf 2002}, pp. 530-538. 
\im
DOUADY S., ANDREOTTI B. et DAERR A., On granular surface flow equations, {\em European Physical Journal }B, 11, {\bf 1999}, pp. 131-142.
\im
POULIQUEN O. et FORTERRE Y., Friction law for dense granular flows. Application of the motion of a mass down a rough inclined plane, 
 {\em Journal of Fluid Mechanics}, 453, {\bf 2002}, pp. 133-151.
\im
MANGENEY A., VILOTTE J-P., BRISTEAU M-O., PERTHAME B., SIMEONI C. et YERNENI S., 
Numerical modelling of avalanches based on Saint Venant equations using a kinetic scheme, 
{\em Journal of Geophysical Research}, in press, {\bf 2003}.
\im
POULIQUEN O., Scaling laws in granular flows down rough inclined planes, {\em Physics of Fluids}, 11, {\bf 1999}, pp. 542-548.
\im
DA CRUZ F., PROCHNOW M., AZANZA E., RAGOUILLIAUX A., TOCQUER L., MOUCHERONT P., ROUX J-N., COUSSOT PH. et CHEVOIR F., 
Ecoulements denses de grains secs sur plan incliné, {\em Actes des Journées Sciences de l'Ingénieur du LCPC}, {\bf 2003}, pp. 541-546.
\en
\end{document}